\documentclass[lettersize,journal]{IEEEtran}
\usepackage{amsmath,amsfonts}
\usepackage{algorithm}
\usepackage{algpseudocode}
\usepackage{array}
\usepackage{textcomp}
\usepackage{stfloats}
\usepackage{url}
\usepackage{verbatim}
\usepackage{graphicx}
\usepackage{cite}
\usepackage{multirow}
\usepackage{xcolor}
\usepackage{bm}
\usepackage{booktabs}

\ifCLASSOPTIONcompsoc
  \usepackage[caption=false,font=normalsize,labelfont=sf,textfont=sf]{subfig}
\else
  \usepackage[caption=false,font=footnotesize]{subfig}
\fi

\begin{document}

\title{Achieving Robust Channel Estimation Neural Networks by Designed Training Data}

\author{\IEEEauthorblockN{Dianxin Luan,~\IEEEmembership{Graduate Student Member,~IEEE,} John Thompson,~\IEEEmembership{Fellow,~IEEE}}\\ 
\IEEEauthorblockA{\textit{Institute for Imaging, Data and Communications, School of Engineering, University of Edinburgh, Edinburgh, EH9 3JL, United Kingdom}}
}

\maketitle

\begin{abstract}
Channel estimation is crucial in wireless communications. However, in many papers neural networks are frequently tested by training and testing on one example channel or similar channels. This is because data-driven methods often degrade on new data which they are not trained on, as they cannot extrapolate their training knowledge. This is despite the fact physical channels are often assumed to be time-variant. However, due to the low latency requirements and limited computing resources, neural networks may not have enough time and computing resources to execute online training to fine-tune the parameters. This motivates us to design offline-trained neural networks that can perform robustly over wireless channels, but without any actual channel information being known at design time. In this paper, we propose design criteria to generate synthetic training datasets for neural networks, which guarantee that after training the resulting networks achieve a certain mean squared error (MSE) on new and previously unseen channels. Therefore, trained neural networks require no prior channel information or parameters update for real-world implementations. Based on the proposed design criteria, we further propose a benchmark design which ensures intelligent operation for different channel profiles. To demonstrate general applicability, we use neural networks with different levels of complexity to show that the generalization achieved appears to be independent of neural network architecture. From simulations, neural networks achieve robust generalization to wireless channels with both fixed channel profiles and variable delay spreads. 
\end{abstract}


\IEEEpeerreviewmaketitle

\section{Introduction}
\IEEEPARstart{F}{or} sixth-generation (6G) wireless communication systems and beyond, the orthogonal frequency division multiplexing (OFDM) is expected to continue as the baseband modulator \cite{wang2023road, wu2024intelligent}. Therefore, an intelligent agent which can obtain precise channel state information is required for better intelligent spectrum sensing and adaptive transmission, particularly in cognitive communications. Least-squares (LS) and minimum mean squared error (MMSE) methods are widely investigated for channel estimation (CE) \cite{van1995channel, edfors1998ofdm}. In addition, a decision-directed iterative algorithm is proposed in \cite{deng2003decision} and the paper \cite{orozco2004channel} performs channel estimation with an implicit sequence added to the data arithmetically. However, conventional approaches cannot meet the current demand for higher data rate and more reliable communications, where artificial intelligence (AI) solutions can provide optimization \cite{neumann2018learning, koller2022asymptotically, baur2024channel, baur2024leveraging, luan2025robust}. Compared with the conventional methods finding closed-form expressions, neural networks aim for an improved and locally optimal solution, such as ChannelNet \cite{soltani2019deep}. With the development of the attention mechanism \cite{bahdanau2014neural}, the papers \cite{mashhadi2021pruning, gao2021attention} exploit the attention methodology explicitly to perform channel estimation for multiple-input multiple-output (MIMO) system. Our prior works \cite{luan2022attention, luan2023channelformer} propose an encoder-decoder architecture to utilize the transformer encoder \cite{vaswani2017attention} to weight the LS inputs according to their relative importance. 

However, data-driven algorithms often degrade on new channels. This instability prohibits the real-world implementation of neural networks for time-variant channels. Although online training methodology can fine-tune the offline-trained neural networks to adapt to new channels, it often requires huge computing resources and time to update neural network parameters. For practical deployment, online training has several critical drawbacks as follows 
\begin{subsubsection}{Latency and computation overhead}
Retraining neural networks online will result in a high processing latency and memory consumption, and it is challenging to collect real-time samples in a very short time. For example, the slot length is usually between 0.0625 milliseconds and 1 millisecond as defined in the 3GPP TS 38.211 document. Therefore, the receiver needs at least 62.5-1,000 milliseconds to collect 1,000 complete sets of channel examples. Moreover, the computational cost and latency of the training process are typically very high, so it is difficult for battery-powered terminals to update neural networks promptly to adapt to new channels, especially in highly dynamic environments. Although online training for dynamic environments has been shown to rapidly converge within several hundred iterations in \cite{luan2023channelformer, zheng2021online} and within 1,500 episodes in \cite{mohajer2024dynamic}, the computing resources required remain significant and unavoidable for this process. 
\end{subsubsection}
\begin{subsubsection}{Performance loss}
The interpolation for the channel estimate is usually challenging \cite{minn2000investigation} because statistical models can be inaccurate, cumbersome and difficult to handle \cite{tang2007pilot, nissel2018doubly}, so the basis expansion model is widely used \cite{borah1999frequency} for doubly-selective channels. Neural networks can simply achieve this by using the complete channel information for training. However, the actual channel state information for the complete packet is difficult to obtain from the noisy received signals. This indicates that neural networks supporting online training will lose the capability of precise interpolation, which is the dominant performance gain as shown in Section.~\ref{Analysis on constructing the training data to train neural networks for wireless channel estimation}. 
\end{subsubsection}

Furthermore, it is impossible for neural networks to learn all the existing channels online. Even if this could happen, neural networks trained with a mixture of channel samples will be susceptible to catastrophic forgetting \cite{mccloskey1989catastrophic} (see Fig.~\ref{Appendix_2}). In that case, they will never perform robustly as the LS estimate and still need online training to compensate for channel mismatches which are very complex for practical implementation. To address these problems, offline-trained neural networks are expected to perform well across different wireless channels, rather than just on the channel used for training. This motivates us to achieve good generalization for channel estimation neural networks, while generalization refers to the model's ability to adapt properly to new and previously unseen data \cite{bousquet2002stability}. 

Compared to online training, neural networks with good generalization do not require any actual channel information of real-world channels nor do they incur additional latency, computation overhead and performance loss arising from online-training during operation. Recent discussions on neural network generalization are provided in \cite{zhang2021understanding, advani2020high, akrout2023domain, wang2020high}. According to \cite{wang2020high}, certain high and low-frequency feature components are shown to affect the generalization properties. For robust channel estimation, channel correlation design helps to resist the channel mismatch for both the MMSE filters \cite{edfors1998ofdm} and the fixed finite impulse response estimator \cite{cavers1991analysis}. A support vector machine (SVM) algorithm for robust channel estimation is proposed in \cite{garcia2006support}, using a complex regression SVM formulation adapted to pilot signals under non-Gaussian impulse noise interference. Reference \cite{cai2004robust} transforms channel estimation problem in OFDM into a set of independent time-domain estimation problems, making it more robust than the Kalman estimation counterpart when dealing with uncertainty. A reconfigurable intelligent surface (RIS) aided channel estimator is proposed in \cite{demir2024efficient} to enhances robustness against spatially correlated electromagnetic interference by optimizing RIS configurations. 

As the channel profile depends on the practical environment where the device operates, it is impossible to know this information in advance, especially when training neural networks offline. This motivates finding neural network solutions which have the necessary reliability for general (not just trained) communication scenarios. As we study channel estimation neural networks, some intrinsic properties of the training data are found to affect their generalization abilities in this paper. 
\subsection{Main contributions}
To resolve this, this paper proposes design criteria for training neural networks to achieve good generalization. This ensures that trained neural networks are able to precisely estimate different channels using the same set of neural network parameters learned from a designed channel, but with no access to channel information of these channels. The proposed method aims to realize good generalization for general but not specific neural networks, thus the impact of neural network architecture on the generalization property is not investigated. The main advantages of the proposed training procedure are 
\begin{itemize}

    \item Regardless of the channels which neural networks operate on, they can achieve a certain mean squared error (MSE) performance by following the proposed design criteria and the BER performance is very close to the theoretical limit (i.e. use the actual and complete channel matrix to recover the received signal). This will provide neural networks with good generalization for real-world implementations, rather than randomly predicting on previously unseen channels. Test results show that eqn.~(\ref{error}) (the MSE error for the designed MMSE filter) provides good indications for actual neural networks. The design criteria is further validated to be effective on more realistic scenarios with varying delay spreads, which comprise a distinct form of channel model. 

    \item Actual and complete channel information can still be used for training because these channels are synthetically generated. Neural networks do not require actual information of test channels when predicting them, but still perform well on these channels. Moreover, by being trained with the complete channel matrix, neural networks still retain precise interpolation capabilities for new, previously unseen channels. This is the dominant performance gain over conventional methods, as shown in Section.~\ref{Analysis on constructing the training data to train neural networks for wireless channel estimation}. 
   
\end{itemize}
Compared to \cite{10279223}, we extend this by providing theoretical analysis of how our design criteria works to achieve such generalization, propose a benchmark design for general scenarios and study the system scalability. We also demonstrate the generalization results by using an extremely low-complexity neural network, which is feasible for state of the art terminals. 
\subsection{Outline of this paper} 
The rest of the paper is organized as follows. Section.~\ref{System architecture and propagation channels} describes the system architecture of the fifth generation New Radio (5G NR) OFDM baseband and the 3GPP TS 36.101 and TR 38.901 channel models. Section.~\ref{Conventional methods and the deployed neural networks} introduces the baseline methods for channel estimation, along with the deployed neural networks. Section.~\ref{Analysis on constructing the training data to train neural networks for wireless channel estimation} proposes the method for designing the training data. Section.~\ref{Simulation results} provides the simulation results. Section.~\ref{Conclusion} summarizes the key findings of the paper. 

\textit{Notations}: In this paper, we use uppercase bold letters for matrices $\mathbf{X}$ and lowercase bold letters for vectors $\mathbf{x}$. The complex Gaussian distribution with mean $\mu$ and variance $\sigma^2$ is written as $\mathcal{CN}(\mu, \ \sigma^{2})$. The conjugate transpose and inverse of matrix $\mathbf{X}$ are represented by $\left(\mathbf{X}\right)^{H}$ and $\left(\mathbf{X}\right)^{-1}$. The function $\mathrm{Tr}\left\{\cdot\right\}$ denotes the trace of a matrix and $\left\Arrowvert\ \cdot\right\Arrowvert_{F}^{2}$ denotes the Frobenius norm of the matrix. $\mathbf{I}_N$ represents an identity matrix of size $N \times N$, $\mathbb{E}$ is the expectation operator, and $\mathrm{diag}\left\{\cdot\right\}$ is the diagonal operator that converts a vector to a diagonal matrix. For scalars, the conjugate operation is denoted by $(\cdot)^{*}$. 
\section{System architecture and propagation channels}
\label{System architecture and propagation channels}
\subsection{Baseband configuration}
\label{Baseband configuration}
\begin{figure*}[htbp]
\centerline{\includegraphics[width=0.57\textwidth]{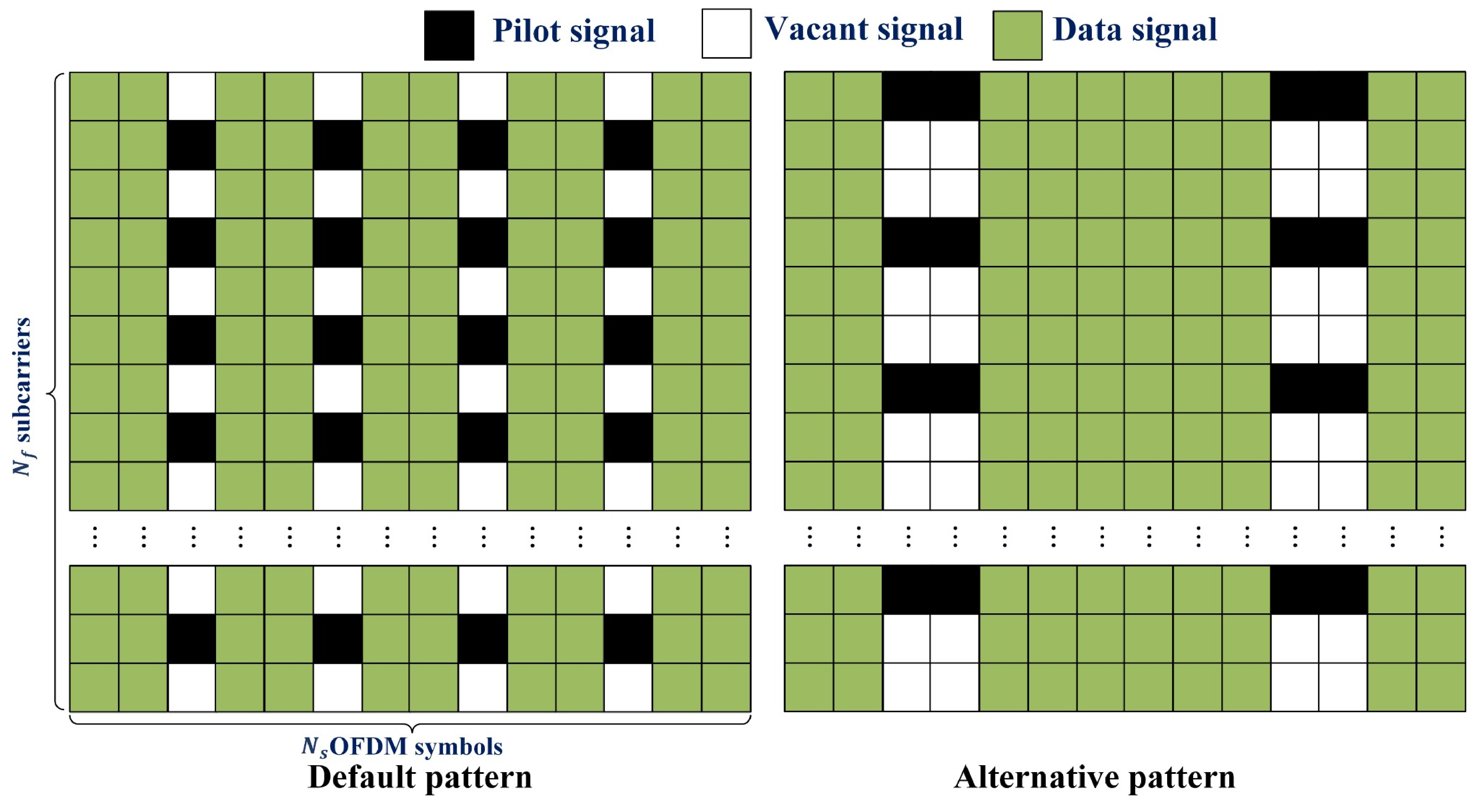}}
\caption{The deployed DM-RS patterns in this paper. The DM-RS additional positions in combination with single-symbol (at left side) and double-symbol DM-RS (at right side)}
\label{DM-RS pattern}
\end{figure*}

This paper considers wireless communication for an OFDM cellular system with a frequency spacing of $f_{space}$ kHz. The source signal $\mathbf{s}$, generated by the uniform distribution randomly, is processed by a Quadrature Phase Shift Keying (QPSK) modulator with Gray coding. The QPSK-modulated symbols are assigned to the data subcarriers as shown in Figure.~\ref{DM-RS pattern}. The pilot signals are known to the receiver, and the vacant pilot subcarriers are set to 0. Each slot consists of $N_f$ subcarriers and $N_s$ OFDM symbols. By following the 5G NR specifications in 3GPP TS 38.211, one of two possible demodulation reference signal (DM-RS) patterns is used:  
\begin{subsubsection}{Default DM-RS pattern}
    A single-symbol DM-RS with three additional positions is deployed as the default pilot pattern. The 3\textsuperscript{rd}, 6\textsuperscript{th}, 9\textsuperscript{th} and 12\textsuperscript{th} OFDM symbols are reserved for pilots ($N_{pilot} = 4$). For each pilot OFDM symbol, the second subcarriers of each indices of $L_s = 2$ subcarriers is reserved for pilot subcarriers. 
\end{subsubsection}
\begin{subsubsection}{Alternative DM-RS pattern}
    The alternative pilot pattern is a double-symbol DM-RS as shown in Fig.~\ref{DM-RS pattern}. The 3\textsuperscript{rd}, 4\textsuperscript{th}, 11\textsuperscript{th} and 12\textsuperscript{th} OFDM symbols are reserved for pilots ($N_{pilot} = 4$). For each pilot OFDM symbol, the first subcarriers of each indices of $L_s = 3$ subcarriers is reserved for pilot. 
\end{subsubsection}

The inverse fast Fourier transform (IFFT) converts the DM-RS frame signal of the slot from the frequency domain to the time domain. Then the Cyclic-Prefix (CP, of length $L_{\mathrm{CP}}$ samples plus an implementation delay of 7 samples) is added to the front of each symbol to resist the multi-path effects. Normally the fast Fourier transform (FFT) and IFFT operators use scaling factors of 1 and $1/{N_f}$, but those are changed to $\left(1/\sqrt{N_f}\right)$ to avoid changing the power of their outputs. The channel is assumed to be a multipath channel with $M$ paths and the corresponding impulse response is 
\begin{equation}
h(\tau, t) = \sum_{m=0}^{M-1} a_m(t)\delta(\tau-\tau_mT_s), 
\label{impulse response}
\end{equation}
where $a_m\left(t\right)$ is the path gain of the $m^{th}$ path, $\tau_m$ is the corresponding delay normalized by the sampling period $T_{s}$. Therefore, the $n$th sampled time-domain $\mathbf{h}(n; t)$ is 
\begin{equation}
    \mathbf{h}(n; t) = \sum_{m=0}^{M-1} a_m\left(t\right)e^{-j\pi\frac{n+(N_f-1)\tau_m}{N_f}}\frac{\sin(\pi\tau_m)}{\sin(\frac{\pi}{N_f}(\tau_m-n))}. 
\label{channel tap}
\end{equation}

The sampled power delay profile (PDP) is given by 
\begin{equation}
    \mathbf{p}(n) = \mathbb{E}\left(\mid \mathbf{h}(n)\mid^2\right). 
\end{equation}

Then, the sampled time-domain received signal $\mathbf{y}$ is given by 
\begin{equation}
\mathbf{y} = \mathbf{h}\otimes \mathbf{x} + \mathbf{n}, 
\label{h}
\end{equation}
where $\otimes$ denotes the convolution operation, $\mathbf{n}$ is assumed to be additive white Gaussian noise (AWGN) for simulation and $\mathbf{x}$ is the transmitted time-domain OFDM signal. Each slot is assigned a new channel realization. After removing the CP, the receiver converts the time-domain data to the frequency-domain using the FFT operation. By assuming that $\forall{\tau_{m}} \leq L_{CP}$, the channel gain at the $k$th subcarrier is 
\begin{equation}
\mathbf{H}\left(k, l\right) \frac{1}{\sqrt{N_f}}\sum_{m=0}^{M-1} a_m\left(T_ol\right)e^{-j2\pi \frac{k\tau_m}{N_f}}. 
\label{Phase}
\end{equation}
\normalsize
for $k = 0, 1,..., N_f-1$ and $T_o$ is one complete OFDM symbol period including the CP. The received frequency domain signal at $k$th subcarrier and $l$th OFDM symbol, $\mathbf{Y}(k, l)$, is of the form: 
\begin{equation}
\mathbf{Y}(k, l) = \mathbf{H}(k, l)\mathbf{X}(k, l) + \mathbf{N}(k, l), 
\end{equation}
where $\mathbf{N} \sim \mathcal{CN}(\mathbf{0}, \ \sigma_{N}^{2}\mathbf{I}_{N_f})$, and complete $\mathbf{H} \in \mathbb{C}^{{N_f}\times N_{s}}$ is the channel matrix to be estimated and $\sigma_X^2 = E\{(N_fN_s)^{-1}\left\Arrowvert\mathbf{X}\right\Arrowvert_{F}^{2}\}$. The received pilot signal is then extracted to provide a reference for the channel matrix of each packet. The recovered signal will be filtered to remove the channel effects, and then processed in a QPSK demodulator to obtain the received bit-level data estimates $\mathbf{\hat{s}}$. 
\subsection{Channel models}
\label{Channel}
This paper considers single-input-single-output downlink scenarios, utilizing the channels with both fixed PDPs and variable delay spreads for simulation. 
\subsubsection{Fixed PDP channels}
They are modeled by using the generalized method of exact Doppler spread method \cite{patzold2009two} operating at carrier frequency $f_{s}$. The extended pedestrian A (EPA), extended vehicular A (EVA) and extended typical urban (ETU) channels are defined in 3GPP 36.101. The customized channel PDPs are given in Table.~\ref{customized channels}. 
\begin{table}[htbp]
\caption{Customized channels with fixed PDP}
\begin{center}
\begin{tabular}{|c|c|}
\hline
\multicolumn{2}{|c|}{Flat fading channel} \\
\hline
Path delay& 0 ms \\
\hline
Average path gain & 0.0 dB \\
\hline
\multicolumn{2}{|c|}{Defined channel 1 (DC1)} \\
\hline
Path delay& [0 0.05 0.1 0.2 0.4] ms \\
\hline
Average path gain & [0.0 -2.0 -4.0 -8.0 -16.0] dB \\
\hline
\multicolumn{2}{|c|}{Defined channel 2 (DC2)} \\
\hline
Path delay& [0 0.03 0.2 0.3 0.5 1.5 2.5 5] ms \\
\hline
Average path gain & [-7.0 0 0 -1.0 -2.0 -1.0 -1.0 -5.5] dB \\
\hline
\multicolumn{2}{|c|}{Defined channel 3 (DC3)} \\
\hline
Path delay& [0 0.05 0.12 0.2 0.23 0.5 1.6 2.3 5 7] ms \\
\hline
Average path gain & [0.0 -1.0 -1.0 -1.0 -1.0 -1.5 -1.5 -1.5 -3.0 -5.0] dB \\
\hline
\multicolumn{2}{|c|}{Two-path fading channel} \\
\hline
Path delay& [0.05 5] ms \\
\hline
Average path gain & [-3.0 -3.0] dB \\
\hline
\multicolumn{2}{|c|}{The designed channel} \\
\hline
Path delay& [0, 0.03, 0.2, 0.3, 0.5, 1.5, 2.5, 5, 7, 9] ms \\
\hline
Average path gain & [0.0, 0.0, 0.0, 0.0, 0.0, 0.0, -1.0, -1.0, -2.0, -3.0] dB \\
\hline
\multicolumn{2}{|c|}{Additional channel 1} \\
\hline
Path delay& [0 0.05 0.12 0.2 0.5 1 1.6] ms \\
\hline
Average path gain & [-7.0 -3.0 -14.0 0.0 -12.0 -5.0 -9.0] dB \\
\hline
\multicolumn{2}{|c|}{Additional channel 2} \\
\hline
Path delay& [0 0.05 0.12 0.2 0.23 0.5 1.6 2.3 5 7] ms \\
\hline
Average path gain & [0.0 -14.0 -15.0 -1.0 -5.0 -1 -15 -10 -10 -5.0] dB \\
\hline
\end{tabular}
\label{customized channels}
\end{center}
\end{table}
\subsubsection{Variable delay spread channels}
Clustered delay line (CDL) and tapped delay line (TDL) channels are modelled as link-level fading scenarios, which represents more realistic channels with carrier frequency of $f_{r}$. The CDL-A, CDL-B, CDL-C, TDL-A and TDL-B channels are implemented following the 3GPP TR 38.901, which have variable root mean square delay spreads ($\mathrm{DS}$). The scaled delays are computed by 
% JST - please define delta Dirac function below eqn (1)
\begin{equation}
\tau_{\mathrm{n, scaled}} = \tau_{\mathrm{n, model}}\mathrm{DS_{desired}}, 
\end{equation}
where $\tau_{\mathrm{n, scaled}}$ is the scaled delay value of $n^{th}$ cluster, $\tau_{\mathrm{n, model}}$ is the corresponding normalized model delay and the $\mathrm{DS_{desired}}$ is the wanted $\mathrm{DS}$. These channels represent more realistic scenarios than the aforementioned channels. 
\section{Conventional methods and neural networks for demonstration}
\label{Conventional methods and the deployed neural networks}
In this section, we will explain the conventional frequency domain estimation algorithms and introduce employed neural networks. 
The LS estimate at the $k$th subcarrier is given by 
\begin{equation}
    \mathbf{\hat{H}_{ls}}(k) = Y_{k}X_{k}^{-1}, 
\end{equation}
where $Y_{k}, X_{k}$ denotes the received and transmitted signals for the $k$th subcarrier. This paper applies bilinear interpolation on $\mathbf{\hat{H}_{ls}^{pilot}} \in \mathbb{C}^{\frac{N_f}{L_{s}}\times N_{pilot}}$ (LS estimate at the pilot positions) to predict the complete channel matrix. 

By weighting the LS estimate to minimize the MSE loss compared to the actual channel matrix, the linear MMSE estimate for OFDM symbols $\mathbf{\hat{H}_{MMSE}} \in \mathbb{C}^{N_f}$ is computed by 
%
% JST - I have put brackets round \frac{\sigma_N^2}\right){\sigma_X^2}
\begin{align}
\label{MMSE}
    &\mathbf{\hat{H}_{MMSE}} = W^{H}\mathbf{\hat{H}_{ls}}, \\    
    &W = \left(\mathbf{R_{HH}} + \left(\frac{\sigma_N^2}{\sigma_X^2}\right) \mathbf{I}_{N_f}\right)^{-1}\mathbf{R_{HH}}, 
\end{align}
where $\mathbf{R_{HH}} \in \mathbb{C}^{{N_f} \times {N_f}}$ is the channel autocorrelation matrix with elements $\mathbf{R_{HH}}(k_1, k_2) = E\{H_{k_1}H_{k_2}^{*}\}$ for $\forall{k_1, k_2 \in [0, N_f - 1]}$. This paper utilizes the actual channel vectors for each pilot symbol $\mathbf{H_{c}} \in \mathbb{C}^{N_f}$ and the actual channel vectors for the corresponding pilot subcarriers of pilot symbol $\mathbf{H_p} \in \mathbb{C}^{\frac{N_f}{L_{s}}}$ to compute the correlation matrix by 
\begin{equation}
\mathbf{R_{H_{c}H_p}} = \mathbb{E}\left\{\mathbf{H_{c}}\mathbf{H_{p}^{H}}\right\}, \mathbf{R_{H_pH_p}} = \mathbb{E}\left\{\mathbf{H_{p}H_{p}^{H}}\right\}. \nonumber
\end{equation}
These matrices are then substituted into Equ.~(\ref{MMSE}) to calculate the $\mathbf{\hat{H}_{MMSE}^{pilot}} \in \mathbb{C}^{N_f}$ for each pilot symbol. Similar to \cite{luan2023channelformer}, bilinear interpolation is applied on the MMSE estimate $\mathbf{\hat{H}_{MMSE}^{pilot}} \in \mathbb{C}^{N_f}$ to compute the channel matrix of the complete slot. By incorporating actual channel knowledge, the MMSE method outperforms the LS estimate, however obtaining actual channel information is difficult in practice. 

Like linear MMSE filters, neural networks take $\mathbf{\hat{H}_{ls}^{pilot}} \in \mathbb{C}^{\frac{N_f}{L_{s}}\times N_{pilot}}$ as the input, but the output $\mathbf{\hat{H}} \in \mathbb{C}^{N_{f} \times N_{s}}$ is an estimate for $\mathbf{H} \in \mathbb{C}^{N_{f} \times N_{s}}$ achieving both time and frequency interpolation. The real and imaginary parts of complex metrics are stored in the third dimension of both the real input and the real output. To demonstrate that the proposed method performs reasonably consistently for different types of neural networks, three neural networks of varying complexity are employed. 
\subsubsection{Channelformer}
This is an attention-based encoder-decoder neural architecture to improve performance by exploiting sub-channel correlations \cite{luan2023channelformer}, which represents high-complexity neural network. This paper refers to the offline Channelformer simply as Channelformer. 
\subsubsection{InterpolateNet}
This is a convolutional neural network with skip connections \cite{luan2021low} and only has 9,442 tunable parameters, which gives an example of medium-complexity neural networks. 
\subsubsection{SimpleNet}
\begin{figure}[htbp]
\centerline{\includegraphics[width=0.5\textwidth]{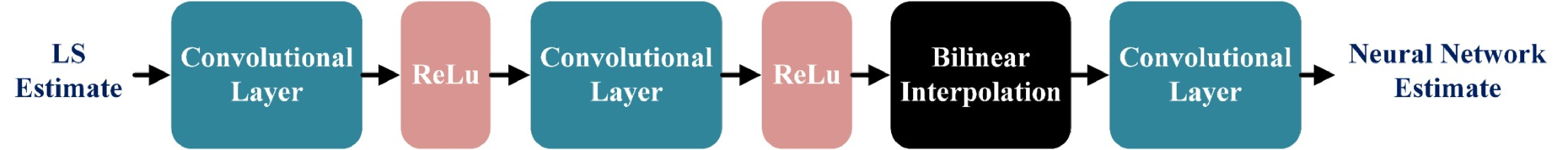}}
\caption{SimpleNet only involves three convolutional layers, two ReLU layers and bilinear interpolation. }
\label{SimpleNet}
\end{figure}

We also use an extremely low-complexity neural network with only \textbf{882 parameters} for demonstration, denoted by SimpleNet as shown in Fig.~\ref{SimpleNet}. The first and the second convolutional layer have 8 filters with a kernel size of [3,3], and the last convolutional layer has 2 filters with a kernel size of [3,3]. Both the critical path length and the number of tunable parameters in the model are even suitable for mobile devices. 
\section{Analysis on constructing the training data for wireless channel estimation neural networks}
\label{Analysis on constructing the training data to train neural networks for wireless channel estimation}
Given that real-world channels are often time-variant, practical neural network implementations should be capable of estimating a wide range of different channels without any prior channel information (similar to the LS estimate), rather than only working on an example channel or similar channels. This indicates that the robust generalization property is essential for real-world neural network implementations. 
\begin{figure}[htbp]
\centerline{\includegraphics[width=0.5\textwidth]{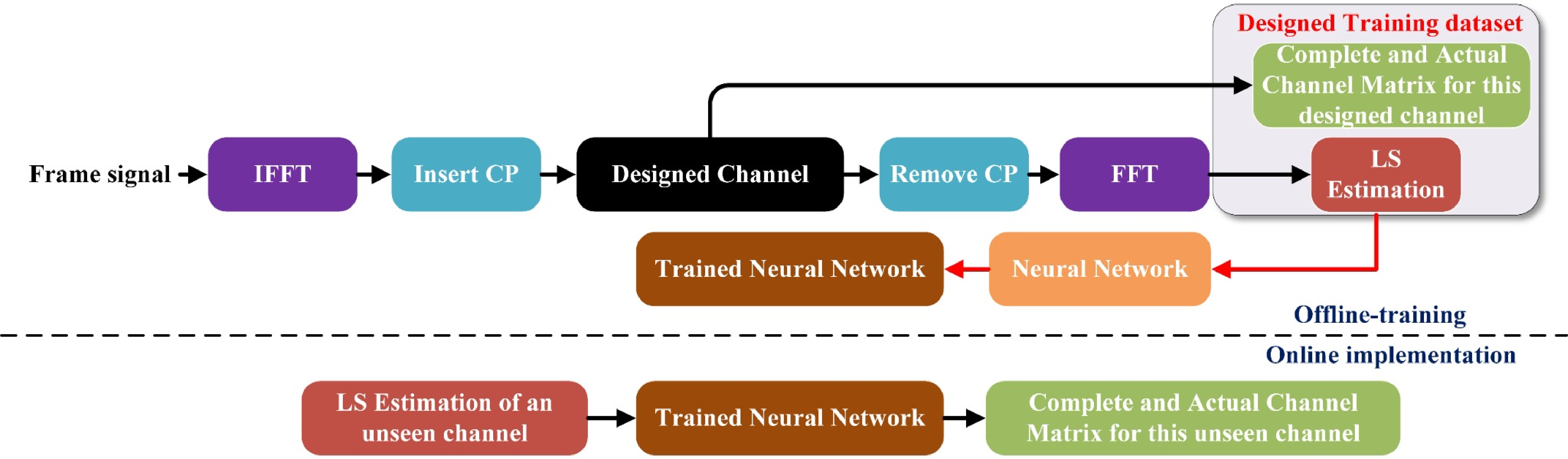}}
\caption{\textcolor{red}{The feature-label pairs of the training dataset are only generated on a designed channel. Offline-trained neural networks can perform well on previously unseen channels, without any actual information of these channels known at design time. }} 
\label{designed training dataset}
\end{figure}

As shown in Fig.~\ref{designed training dataset}, we only train neural networks offline to enable them generalize well over channels. This will ensure neural networks perform robustly even when the physical environment is unseen during the offline training process. As neural networks often perform robustly over Doppler shifts for both low and medium speed scenarios (as shown in Fig. 5b \cite{luan2023channelformer}), the following analysis assumes that $a_m$ is not a function of time. Therefore, this paper primarily addresses the generalization to multi-path fading but Doppler shifts are also modelled when training and testing neural networks. 
\subsection{What makes neural networks outperform other methods}
The channel estimation neural network denoises the LS input and achieves the interpolation mapping: $\mathbb{C}^{\frac{N_f}{L_{s}}\times N_{pilot}} \rightarrow \mathbb{C}^{N_{f} \times N_{s}}$ to predict the complete channel matrix. First, the denoising capability of neural networks only provides a small performance gain as shown in Fig.~\ref{Appendix_1} of Section.~\ref{Simulation results}. As the LS estimate equation shares the same format with the minimum-variance unbiased estimator that attains the Cramer-Rao lower bound \cite{kay1993fundamentals} for estimating the channel gains at the pilot position, the LS estimate error $\epsilon_L$ (compared to the actual channel matrix) is very small, with expected value $E\{\epsilon_L\} = \sigma_{N}^{2}\sigma_X^{-2}$. To further demonstrate that the denoising gain is not significant, neural networks are trained and tested on an example channel but the training label is the actual channel matrix only at the pilot positions, as shown in Fig.~\ref{Appendix_1}. This denoising gain is not comparable to the overall performance gain achieved by neural networks. For example, Channelformer only achieves 10dB gain in Fig.~\ref{Appendix_1} while it achieves over 30dB gain in Fig.~\ref{sub3}. Hence, the denoising capability serves as a secondary factor contributing to neural network performance. \textbf{The overall performance improvement is mainly down to the interpolation capabilities and neural network implementations need to maintain these advantages over wireless channels. }Therefore, neural network solutions are expected to be trained with the actual and complete channel matrix, especially to retain their interpolation capabilities. Due to the fact that actual and complete channel matrix is unavailable in the real world, online-trained neural networks will lose much of their performance gain. This paper only performs offline training of the neural networks using the designed channels to avoid using any actual channel information. Our goal here is to ensure that they generalize well to retain the performance improvement on new and previously unseen channels. 
\subsection{Proposed design criteria to generate training datasets}
\label{Proposed design criteria to achieve robust generalization for channel estimation}
The detailed operation of neural networks is difficult to understand as mathematical interpretability for neural networks is still lacking. Since this paper uses MSE loss, we assume that well-trained neural network’s behaviour is similar to MMSE filters. An alternative expression for the channel autocorrelation matrix is $\mathbf{R_{HH}} = \mathbf{D}\mathbf{\Lambda}\mathbf{D}^{-1}$ \cite{li1998robust, srivastava2004robust} where $\mathbf{D} \in \mathbb{C}^{{N_f} \times {N_f}}$ is the discrete Fourier transform (DFT) matrix and $\mathbf{\Lambda} = \mathrm{diag}\{\mathbf{p}\}, \mathbf{\Lambda} \in \mathbb{R}^{{N_f} \times {N_f}}$. The first $L_{\mathrm{CP}}$ elements of $\mathbf{p}$ contain the main channel power for the deployed OFDM configuration. Then the weighting matrix $\mathbf{W}$ becomes 
\begin{equation}
    \mathbf{W} = \mathbf{D}\left(\mathbf{\Lambda} + \frac{\sigma_N^2}{\sigma_X^2}\mathbf{I}_{N_f}\right)^{-1}\mathbf{\Lambda}\mathbf{D}^{-1}. 
\label{W}
\end{equation}

We now assume that the designed MMSE filter with $\mathbf{\Lambda_{D}} \in \mathbb{R}^{{N_f} \times {N_f}}$ is implemented for a channel with $\mathbf{\Lambda_{A}} \in \mathbb{R}^{{N_f} \times {N_f}}$ and $\mathbf{\Lambda_{D}} \neq \mathbf{\Lambda_{A}}$, then the estimate error $\epsilon_{M} = \mathbb{E}\left\{\mid \mathbf{\hat{H}_{MMSE}} - \mathbf{H}\mid_{2}^2\right\}$ caused by this mismatch is given by 
\begin{equation}
    \epsilon_{M} = \frac{1}{N_f}\mathrm{Tr}\left\{\mathbf{D}\left(\mathbf{I}_{N_f}+\frac{\sigma_N^2}{\sigma_X^2}\mathbf{ZZ}-\mathbf{\Lambda_{A}}\left(2\mathbf{Z} - \mathbf{ZZ}\right)\right)\mathbf{D}^{-1}\right\}, 
\label{error}
\end{equation}
where $\mathbf{Z} = \left(\mathbf{\Lambda_{D}} + \frac{\sigma_N^2}{\sigma_X^2}\mathbf{I}_{N_f}\right)\mathbf{\Lambda_{D}}, \mathbf{Z} \in \mathbb{C}^{{N_f} \times {N_f}}$. From the expression of this error shown in Equ.~(\ref{error}), the mismatch quantitative between that for $\mathbf{\Lambda_{D}} \in \mathbb{R}^{{N_f} \times {N_f}}$ and $\mathbf{\Lambda_{A}} \in \mathbb{R}^{{N_f} \times {N_f}}$ is determined by 
\begin{equation}
    \xi_{M} = \frac{1}{N_f}\mathrm{Tr}\left\{\mathbf{D}\mathbf{\Lambda_{A}}\left(2\mathbf{Z} - \mathbf{ZZ}\right)\mathbf{D}^{-1}\right\}. 
\label{Q}
\end{equation}

This indicates that the MSE of MMSE filters is determined by the PDP of both the designed channel and the actual channel being measured. For channel estimation neural networks, we find that by being trained on a channel with high-eigenvalue and high-rank autocorrelation (by setting a higher delay spread and path gain for PDP), trained neural networks will generalize well to different wireless channels. 

Based on the channel description in Equ.~(\ref{Phase}) and the statistical independence of different channel paths assumption in Equ.~(\ref{assumption}), the $k_1$th column and $k_2$th row element of the channel autocorrelation matrix $\mathbf{R_{HH}}(k_1, k_2)$ is given by 
\begin{equation}
\mathbf{R_{HH}}(k_1, k_2) = \mathbb{E}\left\{\sum_{m=0}^{M-1}A_me^{\frac{-j2\pi\tau_m\left(k_1-k_2\right)}{N_f}}\right\}, 
\label{correlation matirx}
\end{equation}
\begin{equation}
\forall m, \ m': \ \mathbb{E}\left\{a_ma_{m^{'}}^*\right\} = \begin{cases}
    A_m & \mathrm{if} \ m=m' \\
    0 & \mathrm{if} \ m \neq m' \\
\end{cases}, 
\label{assumption}
\end{equation}
where $A_m = \mathbb{E}\left\{\left| a_m \right|^{2}\right\}$ is the average path power. The eigenvalue decomposition of $\mathbf{R_{HH}} \in \mathbb{C}^{{N_f} \times {N_f}}$ for general cases is given by 
\begin{equation}
\mathbf{R_{HH}} = \mathbf{U\Lambda U}^{H}, 
\label{equation}
\end{equation}
where $\mathbf{U} \in \mathbb{C}^{{N_f} \times {N_f}}$ contains the eigenvectors. Based on Equ.~(\ref{W}), the MMSE estimate becomes 
\begin{equation}
\begin{aligned}
\mathbf{\hat{H}} = \mathbf{U}\mathrm{diag}\{\bm{\delta}\}\mathbf{U}^{H}\mathbf{\hat{H}_{ls}}; \\ 
\forall{k \in \left[1,N_f\right]}, \ \bm{\delta}(k) = \frac{\mathbf{p_{A}}(k)}{\mathbf{p_{D}}(k) + \frac{\sigma_N^2}{\sigma_X^2}}. 
\label{MMSE_QPSK}
\end{aligned}
\end{equation}

The analysis above, especially Equ.~(\ref{Q}) and Equ.~(\ref{MMSE_QPSK}), show that a proper design for the PDP of the designed channel may help improve the generalization property of channel estimation neural networks. For the MSE performance of neural networks, Equ.~(\ref{Q}) can be further simplified to 
\begin{equation}
\xi_{M} = \frac{1}{N_f}\sum_{k=0}^{L-1} \mathbf{p}_A(k)\left(2\mathbf{Z}(k, k)-\left(\mathbf{Z}(k, k)\right)^2\right), 
\label{Q_2}
\end{equation}
where $L$ is the minimum length of the designed channel $L_D$ and the length of the implemented channel $L_A$. As the sampled PDP of the actual channel being measured $\mathbf{p}_A \in \mathbb{R}^{{N_f}}$ is determined by physical environments, it is still possible to design a proper channel profile $\mathbf{p}_D \in \mathbb{R}^{{N_f}}$ for $\mathbf{Z} \in \mathbb{C}^{{N_f} \times {N_f}}$ to generate training samples, and ensure trained neural networks achieve a certain MSE for general channels. From the structure of $\mathbf{\Lambda}$, the $\mathbf{R_{HH}}$ of this designed channel is expected to contain a higher number and amplitude of main eigenvalues than those of applicable channels. Applicable channels refer to new and previously unseen channels that trained neural networks can generalize to. As each path's phase follows a uniform distribution over $[-\pi, \pi]$ and is uncorrelated to its power, this paper investigates the impact of path gains and the corresponding delays (PDP) to create the desired channel when generating the training samples for neural networks. As shown in Equ.~(\ref{channel tap}) above, non-integer delays cause leakage effects but these leakage patterns are typically quite localised in the impulse response. Therefore, the designed channel is expected to have a sampled PDP to match most of the expected delays in the applicable channels. However, it also notes that neural network behaviors do not completely align with MMSE formats. 

To create a desired designed channel, we assume that the designed channel $h_{D}$ is a Rayleigh fading channel modelled by \cite{dent1993jakes} which has a fixed PDP that starts from 0dB at the 0ns with specified power values at different delays. The PDP of the designed channel $P_{D}$ is 
\begin{equation}
P_{D}(t) = \sum_{i=0}^{N_{D} - 1} \mathbf{\theta_{D}}\left(i\right)\delta\left(t - \mathbf{\tau_{D}}\left(i\right)\right), 
\label{designed_pdp}
\end{equation}
where $\delta(t)$ denotes the impulse function, $\mathbf{\tau_{D}}\left(i\right)$ is the delay of the $i$th path, $\mathbf{\theta_{D}}\left(i\right)$ is the corresponding path gain and $N_{D}$ is the number of independent paths. These paths correspond to a set of $L_{\mathrm{CP}}$ orthogonal exponential basis functions. Therefore, our method is applicable to any channel whose maximum delay is less than the time duration of the CP. The corresponding coefficients are what neural networks learn from the training procedure. Our approach is based on the ideal MMSE analysis to ensure the neural networks recognize these parameters for generalized estimation. However, the Cramer Rao lower bound (CRLB) always increases when estimating more parameters \cite{kay1993fundamentals}, especially for more delay taps. This indicates that if we have a PDP with a larger number of paths or taps, its estimation performance will be poorer than that for a channel with fewer paths or taps. This aligns with Equ.~(\ref{Q}) where $\xi_M$ will increase with a larger and longer $\mathbf{\Lambda_{D}}$ for arbitrary $\mathbf{\Lambda_{A}}$. It is worth noting that, the primary aim of the proposed approach is that using designed channels to train neural networks can offer robust generalization to new and previously unseen channels. 

\begin{figure*}[htbp]
\centering
\subfloat[Example of PDP design ($h_{D}$) for the applicable channels $h_{A_{0}}$, $h_{A_{1}}$, $h_{A_{2}}$, $h_{A_{3}}$, $h_{A_{4}}$. \label{hT}]{%
       \includegraphics[width=0.5\linewidth]{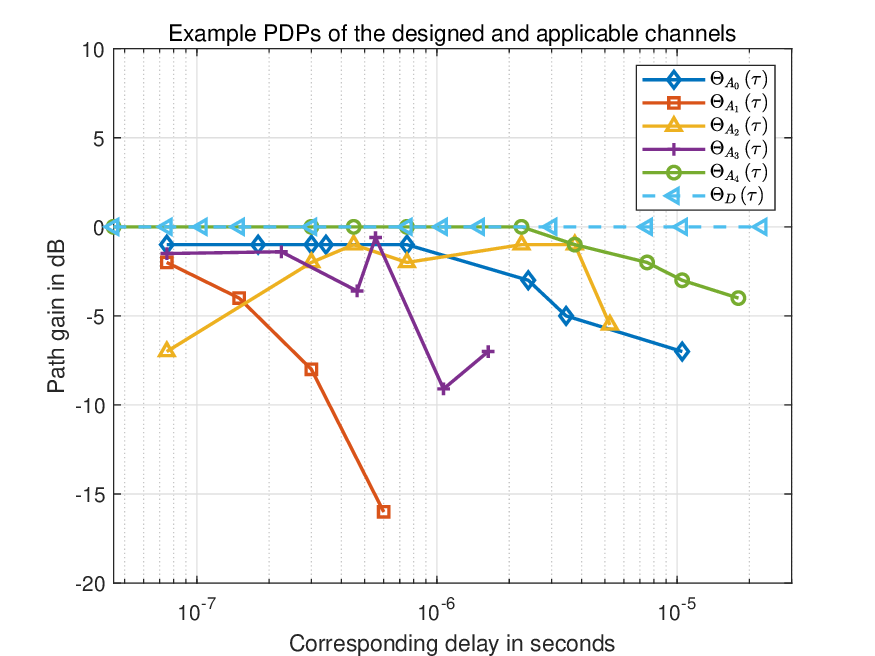}}
\hfill
\subfloat[The magnitude values of channel taps for $h_{D}$, $h_{A_{0}}$, $h_{A_{1}}$, $h_{A_{2}}$, $h_{A_{3}}$, $h_{A_{4}}$ channels are calculated by Equ.~(\ref{channel tap}) \label{channel_taps}]{%
        \includegraphics[width=0.5\linewidth]{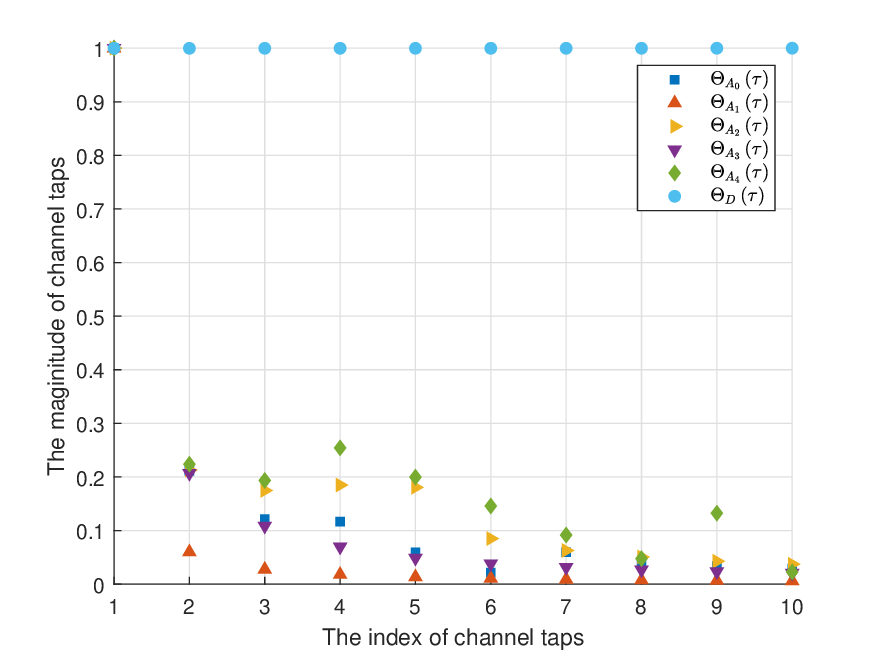}}
\caption{The PDP design for the applicable channels is shown on the left side. \textcolor{red}{It is highlighted that the applicable channels include but are not limited to $h_{A_{0}}$, $h_{A_{1}}$, $h_{A_{2}}$, $h_{A_{3}}$, $h_{A_{4}}$.} Intuitively, channels designed for training neural networks often exhibit high variance in their taps across the applicable channels (we normalized the channel taps by using the maximum magnitude of taps for each channel and only plot the first $L_{\mathrm{CP}}$ taps)}
\end{figure*}

We now consider the applicable channel $h_{A}$ (neural networks trained on $h_{D}$ can generalize to this channel). The corresponding PDP $P_{A}(t)$ is assumed to be 
\begin{equation}
P_{A}(t) = \sum_{i=0}^{N_{A} - 1} \mathbf{\theta_{A}}\left(j\right) \delta\left(t - \mathbf{\tau_{A}}\left(j\right)\right), 
\label{applicable_PDP}
\end{equation}
where $N_{A}$ denotes the number of channel paths for the applicable channels, $\mathbf{\tau_{A}}\left(j\right)$ is the path delay and $\mathbf{\theta_{A}}\left(j\right)$ denotes the average path gain. As shown in Fig.~\ref{hT}, if a neural network is trained on a designed channel $h_{D}$, the applicable channels described in (\ref{applicable_PDP}) need to satisfy the conditions: 
\begin{enumerate}
\item[(C1)] $\forall \tau \in \left[\tau_{A}\left(0\right), \tau_{A}\left(N_{A} - 1\right)\right], \ \Theta_{A}\left(\tau\right) \leq \Theta_{D}\left(\tau\right)$; 
\item[(C2)] $\tau_{A}\left(N_{A} - 1\right) \leq \tau_{D}\left(N_{D} - 1\right), \ \ N_{A} \leq N_{D}.$ 
\end{enumerate}

The continuous function $\Theta_{D}\left(\tau\right)$ of the $h_{D}$ is obtained by linear interpolation between each two adjacent PDP's power-delay pairs. $\Theta_{A}\left(\tau\right)$ is calculated identically as $\Theta_{D}\left(\tau\right)$. Condition (C1) ensures that the power for each delay path of the designed channel is the largest among the applicable channels. Similarly, condition (C2) ensures that the number of paths and maximum path delay of the designed channel is the largest among all possible applicable channels. The number of samples $L_{\mathrm{CP}}$ in the cyclic prefix can be considered as the maximum expected number of taps in the multipath channel or the rank of the channel autocorrelation matrix. Therefore, a practical choice for the designed channel is to ensure the power delay profile covers all delays in the range from delay 0 to delay $T_{CP}$. 
The proposed method is summarized in Algorithm.~\ref{proposed method} below. 
\begin{algorithm}[H]
\caption{Generate Designed Training Data}
\label{proposed method}
\begin{algorithmic}[1]
\Require PDP parameters $(\mathbf{\theta_D, \tau_D})$, number of paths $N_D$, cyclic prefix length $L_{CP}$. 
\Ensure Training dataset $D = \{(H_{ls}^{(i)}, H^{(i)})\}_{i=1}^{N}. $ 

\Ensure (C1) $\forall \tau \in [\min(\tau_A), \max(\tau_A)], \Theta_A(\tau) \leq \Theta_D(\tau). $ 
\Ensure (C2) $\mathbf{\tau_{A}}\left(N_{A} - 1\right) \leq \mathbf{\tau_{D}}\left(N_{D} - 1\right), \ \ N_{A} \leq N_{D}.$ 

\State \textbf{Define designed PDP:}
\For{$i = 0$ to $N_D - 1$}
    \State Assign $\mathbf{\tau_D}(i)$ and $\mathbf{\theta_D}(i)$ to generate channel $\mathbf{h_D}$. 
\EndFor

\State Initialize dataset $D \gets \emptyset$ 
\For{each sample $s$ in the $D$} 
    \State Compute complete channel matrix $\mathbf{H}$ using $\mathbf{h_D}$ and the OFDM system model, 
    \State Collect LS estimate $\mathbf{H_{ls}}$ at the receiver, 
    \State Form feature-label pair $(\mathbf{H_{ls}, H})$ and append it to dataset $D$. 
\EndFor

\State Train neural networks using dataset $D$. 
\end{algorithmic}
\end{algorithm}

Based on the proposed design criteria, we thus propose a channel design for general scenarios, which is denoted as the CE channel. It has the path delay sampled from 0ms to the duration of the CP and each averaged path gain is set to 0dB correspondingly. By being trained on this CE channel, the trained neural networks will generalize well over wireless channels because the received power is normalized to no more than 0dB. Equ.~(\ref{CE}) provides the PDP of the CE channel with a sampling period of $\zeta$ to give $\lfloor \frac{T_{CP}}{\zeta} \rfloor$ paths. 
\begin{equation}
\theta_{CE}(\tau) = 10\mathrm{log_{10}}\left(\sum_{i=0}^{\lfloor \frac{T_{CP}}{\zeta} \rfloor - 1} \delta\left(\tau-\zeta i\right)\right), 
\label{CE}
\end{equation}
where the averaged path gain $\theta_{CE}$ is in dB. According to the design criteria aforementioned, this design should provide the maximum generalization. However, if it is known that $h_A$ channels have a smaller maximum delay or power, the neural network performance will be improved by being trained on another $h_D$ whose maximum delay or power slightly exceeds those of $h_A$ channels. On another hand, flat PDP design is also consistent with the conclusion proposed in \cite{srivastava2004robust}. In this case, all the elements of $\mathbf{Z} \in \mathbb{C}^{{N_f} \times {N_f}}$ are equal (use $z$ to denote) and reconsider Equ.~(\ref{Q_2}) we have 
\begin{equation}
\xi_{M} = \frac{2z-z^2}{N_f}\sum_{k=0}^{L-1} \mathbf{P_A}(k). 
\end{equation}

The $h_D$ should be long enough to have $L_A \leq L_D$. This will ensure $\sum_{k=0}^{L_D-1} \mathbf{P_A}(k) = \sum_{k=0}^{L_A-1} \mathbf{P_A}(k)$ is a constant for arbitrary $h_A$, which indicates that neural network could achieve a certain MSE for different $h_A$. Otherwise, $\sum_{k=0}^{L_D-1} \mathbf{P_A}(k)$ is not a constant because some channel power is localized outside of $L_D$, thus this is not an applicable channel. This design for MMSE filters provides approximately 5dB gain over LS estimate as shown in \cite{luan2023channelformer}, but neural networks trained on this designed channel significantly outperform LS estimate. 

For non-Gaussian noise, the analysis remains valid because in the derivation of the matrix ${\sigma_N}^{2}\mathbf{I}_{N_f}$, which is contained in $\mathbf{Z}$ of Equ.~(\ref{Q}), only requires the noise elements to be independent of each other, i.e. $\mathbb{E}\left(\mathbf{N}\mathbf{N}^{H}\right) = {\sigma_N}^{2} \mathbf{I}_{N_f}$. Therefore, this applies to any additive independent noise source with zero mean and finite power. 

One key benefit of this proposed solution is that, trained neural networks will be able to precisely estimate the channels that are previously unseen from the training dataset, rather than randomly predict on these channels. In this way, trained neural networks are still able to precisely estimate the complete channel matrix on these channels, retaining the interpolation capability which is the main performance gain. 

Another advantage is that, channel estimation neural networks will only require offline-training for practical implementation. The complexity of this proposed method is not important because these procedures are conducted offline. Both the training samples and duration for offline training are allowed to be any desired size, to obtain well-trained neural network parameters for online deployment. An improved performance can be easily attained by setting both the training epoch and the training dataset size to larger values than those used in this paper. As regularization prevents overfitting as well as to improve generalization slightly, the L2 regularization is set to be $0$ to avoid that positive effect in this paper. This is to show generalization is due to the channel used for training neural networks. \textbf{Moreover, the proposed method is applicable for most neural networks because no specific constraints on the neural architecture are imposed and the discussion on how neural network architecture design affects generalization remains independent of the proposed method in this paper. }
\section{Simulation results}
\label{Simulation results}
\begin{figure*}[htbp]
\centering
\subfloat[Denoising performance\label{Appendix_1}]{%
       \includegraphics[width=0.35\linewidth]{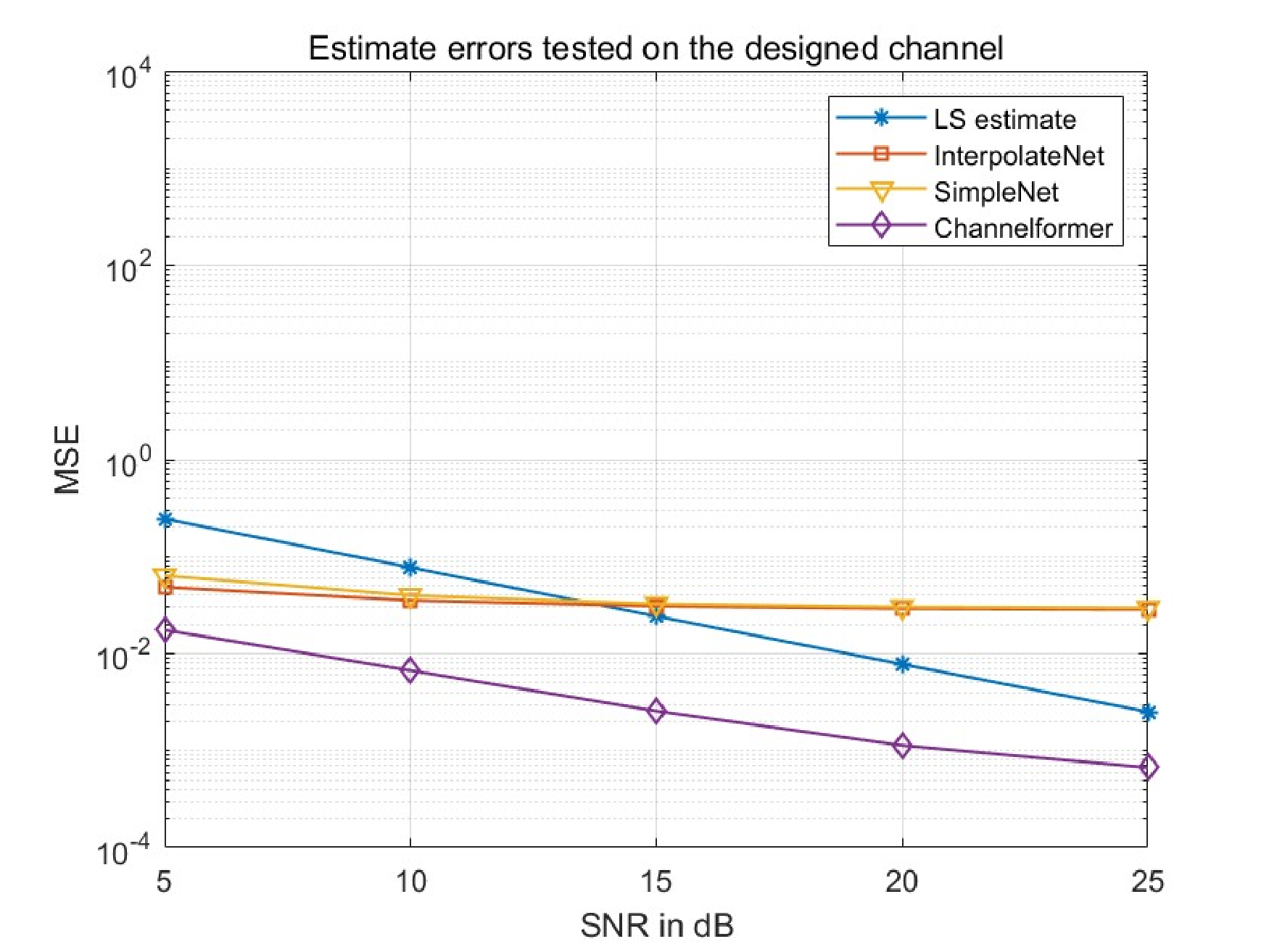}}
\subfloat[Catastrophic interference\label{Appendix_2}]{%
        \includegraphics[width=0.35\linewidth]{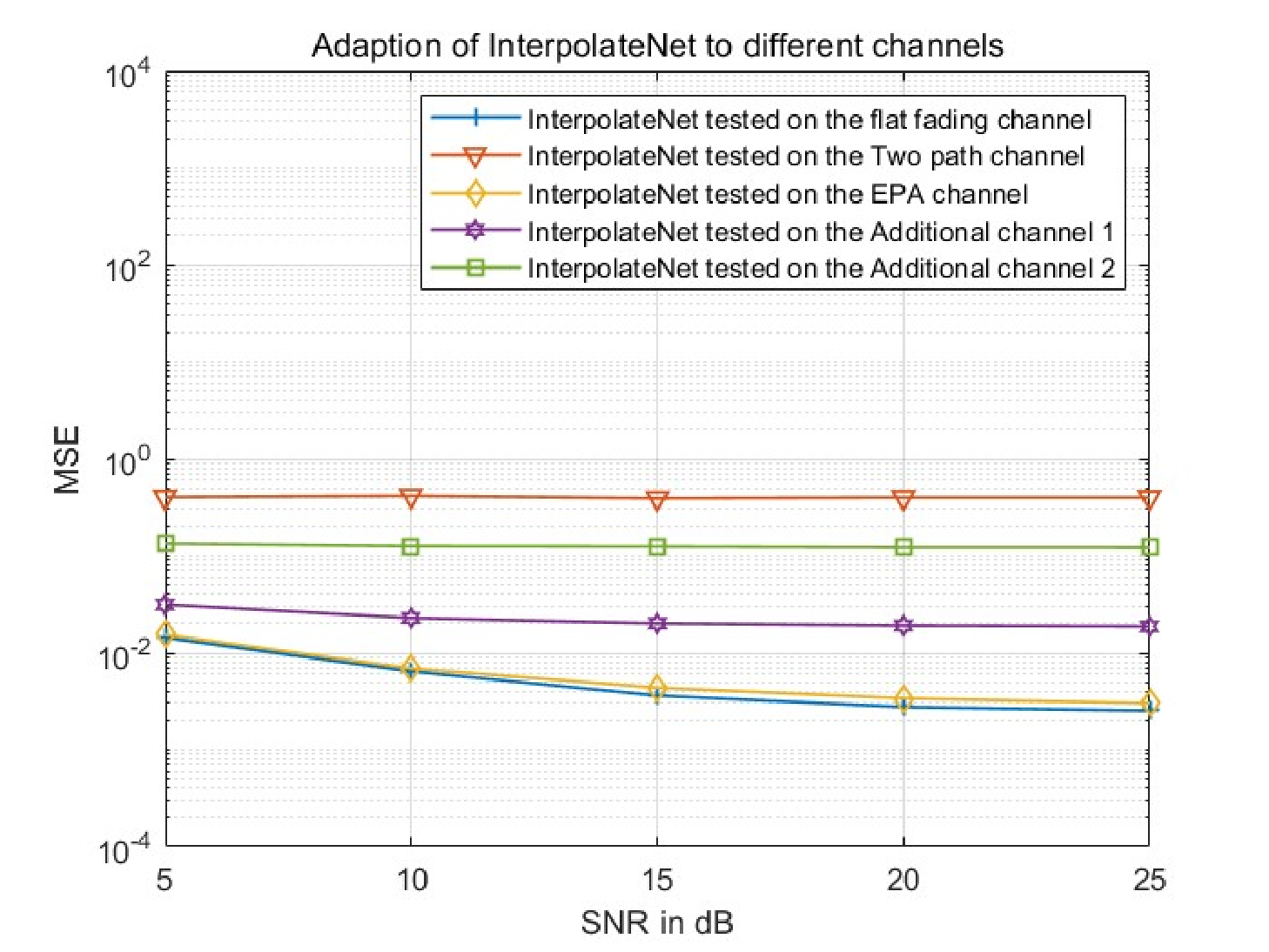}}
\subfloat[Theoretical BER results with perfect channel state information \label{BER}]{%
        \includegraphics[width=0.35\textwidth]{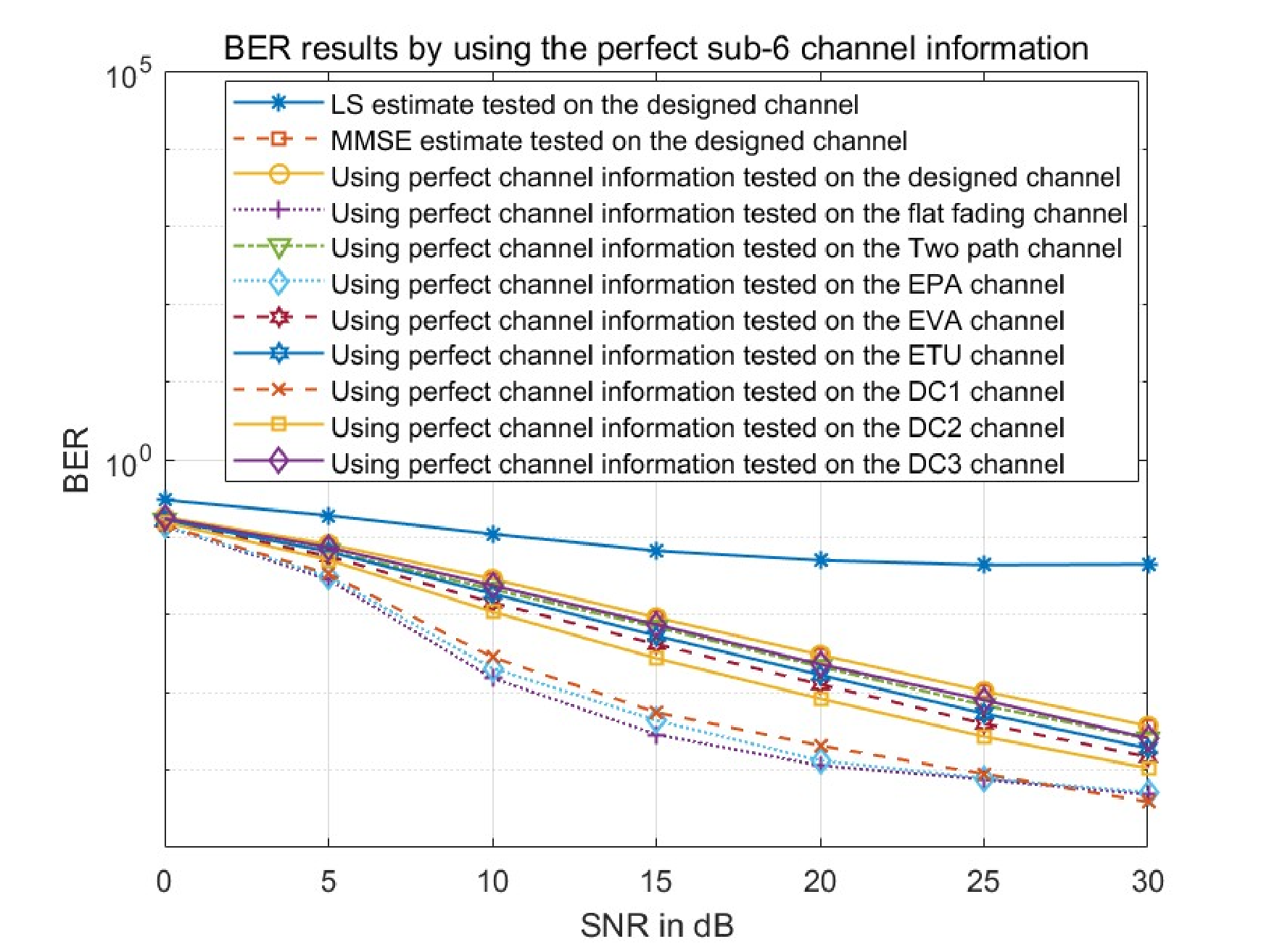}}
\caption{InterpolateNet trained with the actual channel matrix at the pilot positions (left side), trained with mixture of a wide range of channels (middle side) and the BER results with perfect channel state information tested on the fixed PDP channels (right side)}
\end{figure*}

This section provides the simulation results for each deployed neural network. MSE is a key performance metric that evaluates the distance between the actual channel and the estimate for all resource elements in the slot, defined as 
\begin{equation}
    \mathrm{MSE}\left(\mathbf{\hat{H}}, \mathbf{H}\right) = (N_fN_s)^{-1}\mathbb{E}\left\{\left\Arrowvert\mathbf{\hat{H}} - \mathbf{H}\right\Arrowvert_{F}^{2}\right\}, 
\end{equation}
where $\mathbf{H} \in \mathbb{C}^{{N_f} \times {N_s}}$ is the actual channel matrix, $\mathbf{\hat{H}} \in \mathbb{C}^{{N_f} \times {N_s}}$ is the corresponding estimate. The BER is another metric defined as the ratio of mismatch between the $\mathbf{\hat{s}}$ and $\mathbf{s}$, which provides an insight on the precision at the data positions. The value $\Delta$SNR quantifies the signal-to-noise ratio (SNR) difference between that required by the actual channel matrix and that required by the neural network estimate to achieve a certain BER. Fig.~\ref{BER} provides the BER results when the channel matrix is known perfectly, which is the theoretical lower bound. The training dataset consists of 125,000 independent Rayleigh faded channel samples (95\% for training and 5\% for validation) generated following \cite{dent1993jakes}. The SNR range is from 5dB to 25dB and the maximum Doppler shift is from 0Hz to 97Hz. The loss function for the InterpolateNet and SimpleNet is the MSE loss, and for the Channelformer is the Huber loss given in \cite{luan2023channelformer}. The hyper-parameters are shown in Table.~\ref{system hyper-parameters}. The drop factors are 0.5 for Channelformer/SimpleNet and 0.2 for InterpolateNet respectively. To average out Monte Carlo effects, each sample of the simulation curves is tested with 5,000 independent channel realizations. 

Fig.~\ref{Appendix_2} shows neural network performance when being trained with an equal mixture of samples of these five channels. The trained InterpolateNet is shown to perform poorly on some channels that have being already trained with. For example, the trained InterpolateNet cannot perform on the Two-path channel and these MSE values spread from 0.03 to 0.39 at 20dB SNR for these channels. It indicates that online training will not help neural networks to understand or memorize all the channels have already being trained on so this procedure cannot be shut down for practical deployment. This disadvantage and the importance of interpolation capability aforementioned reflect the importance of our proposal. 
\begin{table*}[htbp]
\caption{Simulation hyper-parameters}
\begin{center}
\begin{tabular}{|c|c|c|c|c|c|}
\hline
\multicolumn{2}{|c|}{Offline-training hyper-parameters} & \multicolumn{2}{c|}{System parameters (fixed-PDP channels)} & \multicolumn{2}{c|}{System parameters (CDL/TDL channels)}\\
\hline
Optimizer& Adam& $N_f$& 72& $N_f$& 72\\
\hline
Maximum epoch& 100& $N_s$& 14& $N_s$& 14\\
\hline
Initial learning rate (lr) & 0.002& $L_{\mathrm{CP}}$& 10& $L_{\mathrm{CP}}$& 10\\
\hline
Drop period for lr& 20& $f_{s}$& 2.1GHz& $f_{r}$& 39GHz\\
\hline
Drop factor for lr& 0.5 (0.2)& Maximum Doppler shift & 97Hz& Maximum Doppler shift& 97Hz\\
\hline
Minibatch size& 128& $f_{space}$& 15kHz& $f_{space}$& 15kHz\\
\hline
L2 regularization& 0& $T_{s}$& 9.3897e-07s& $T_{s}$& 9.3897e-07s\\
\hline
\end{tabular}
\label{system hyper-parameters}
\end{center}
\end{table*}
\subsection{Ablation experiments: Validation of the design criteria}
\label{Simulation_ablation}
This section validates the proposed design criteria by the ablation experiment. We apply the proposed design criteria for training, to show whether the trained neural network exhibits strong generalization. Simultaneously, we intentionally violate this design criteria for training to observe the degradation in MSE performance, and this ablation experiment is used to understand the contribution of our proposal. Based on the proposed analysis, the neural network trained on the channels with larger eigenvalues should generalize well to those with lower magnitudes. The generalization relationship is illustrated in Fig.~6d above. Neural networks are trained on the flat fading, EPA, ETU, DC3 and designed channels and then tested on these channels with SNR of 10dB to validate the proposed method. For example, this designed channel is designed to have the largest eigenvalues among all these channels so the neural networks trained on this designed channel should have almost same MSE performance on these channels. In addition, the neural network trained on the DC3 channel will maintain nearly identical performance on the flat fading, EPA, ETU but degrades on the designed channels. 
\begin{figure*}[htbp]
\centering
    \subfloat[InterpolateNet]{\includegraphics[width=0.5\linewidth]{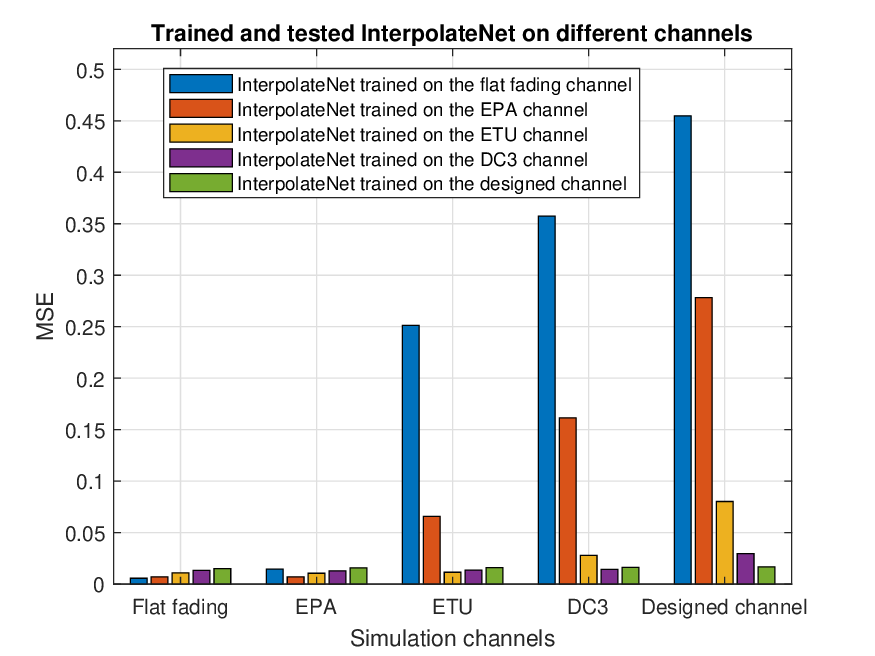}}\hfill 
    \subfloat[Channelformer]{\includegraphics[width=0.5\linewidth]{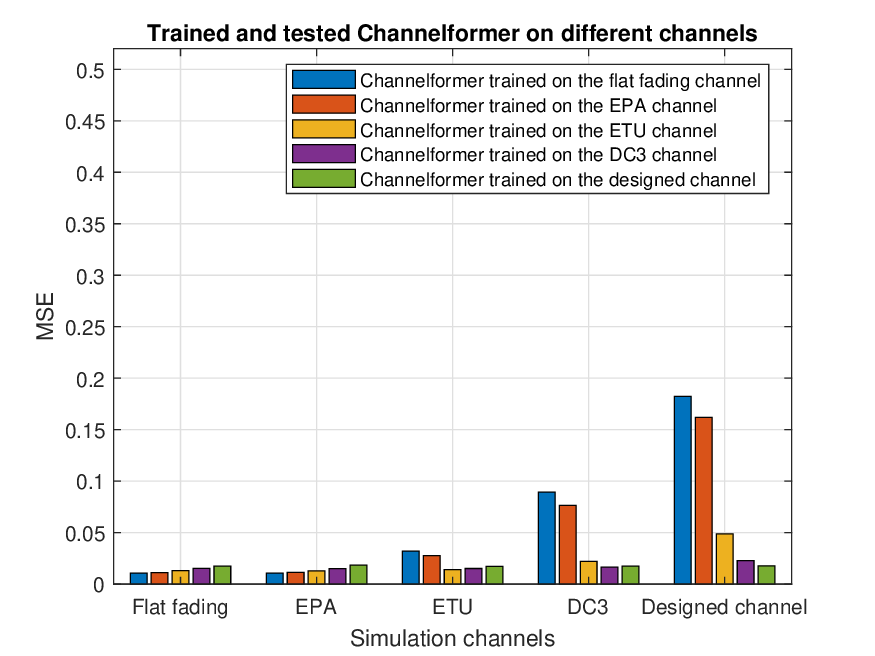}}\hfill 
    \subfloat[SimpleNet]{\includegraphics[width=0.5\linewidth]{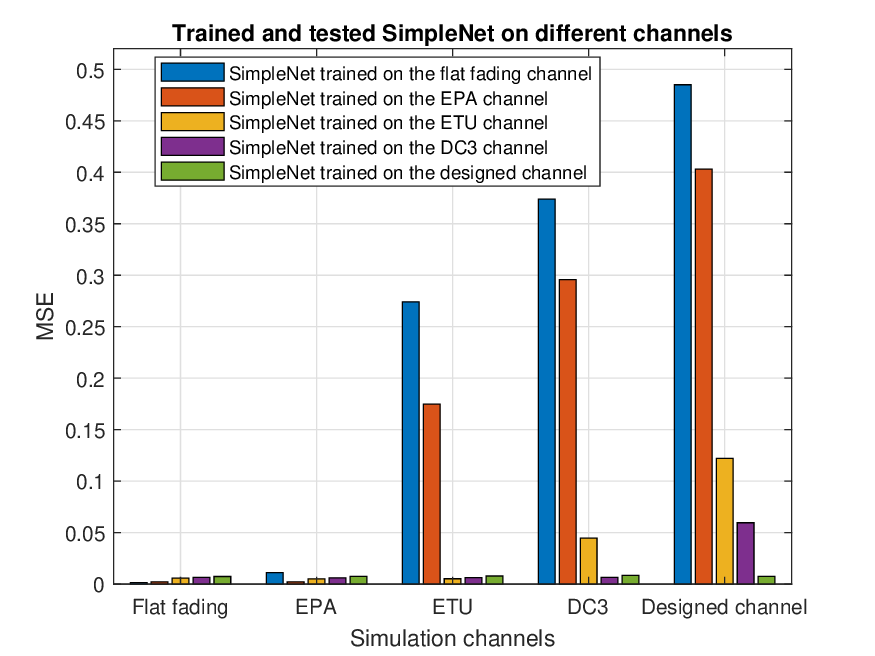}}\hfill 
    \subfloat[MMSE filter]{\includegraphics[width=0.5\linewidth]{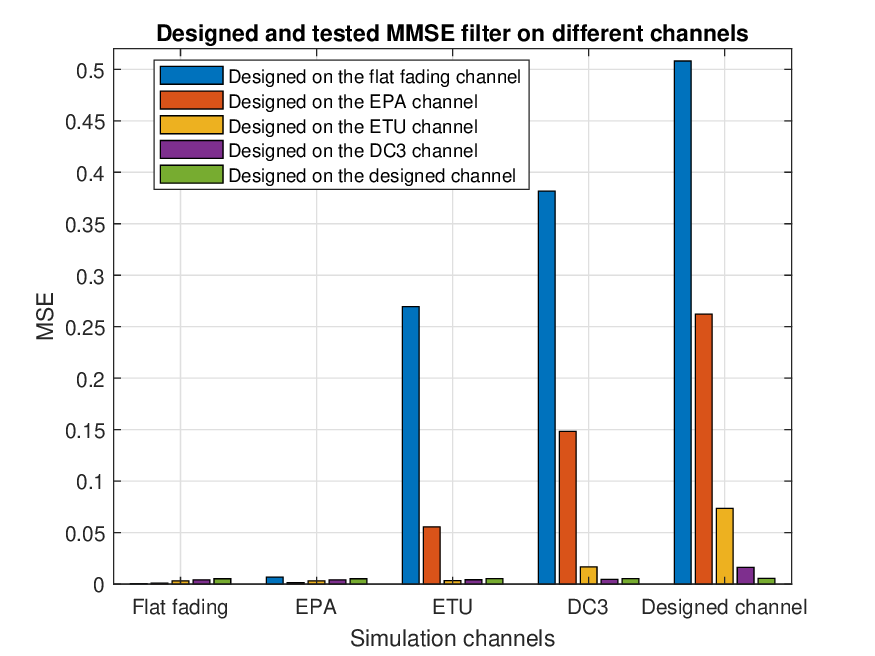}}\hfill 
    \caption{MSE values for the trained InterpolateNet, SimpleNet and Channelformer tested on the flat fading, EPA, ETU and designed channels is shown in Fig.~6a, 6b, 6c. The MSE performance of the MMSE filters designed on different channel profiles using (C1) and (C2) is shown in Fig.~6d. }
\end{figure*}

Fig.~6a, 6b, 6c provide the simulation results when the neural networks are trained and tested on different channels for the ablation experiment. The designed channel trained neural networks generalize well to the ETU, EPA and flat fading channels with an almost same MSE of 0.0158 for InterpolateNet, 0.0075 for Channelformer and 0.0175 for SimpleNet. For the neural networks trained on the DC3 channel, they maintain nearly same performance on the EPA, ETU and flat fading channel while the MSE values increase from 0.0142 to 0.0296 for InterpolateNet, from 0.0065 to 0.0596 for Channelformer and from 0.0162 to 0.0228 for SimpleNet on the designed channel. For neural networks trained on the ETU channel, they maintain nearly the same performance on the EPA and flat fading channel and cannot generalize to the DC3 and designed channel. The EPA trained neural networks only generalize to the flat fading channel. The neural networks trained on the flat fading channel badly degrade when testing on other channels, but have the best performance on this flat fading channel. It should be noted that Fig.~6d only evaluates the MSE for pilot signals (not for the complete packet), so MSE values can be slightly lower than those of the neural networks. However, the trends in Fig.~6a-6c match well in general with Fig.~6d. In conclusion, the simulation results match the hypothesis proposed in this paper. Moreover, it also shows that the performance of a same neural network can be easily improved by using a simple channel to train and test, but this exceptional improvement in MSE performance does not generalize to other channels. It is a crucial point that the designed channel has a worse MSE performance for flat fading than when we train the neural networks only for the flat fading channel, which is the "price" of generalization property. But these results are quite normal as shown in Equ.~(\ref{error}) that MSE will always decrease if the $\mathbf{\Lambda_{D}}$ has a smaller magnitude. 
\subsection{Generalization to fixed PDP channels}
\label{Simulation_sub6}
To show the MSE performance on different fixed PDP channels, the offline-trained neural networks are tested on these channels described in Section.~\ref{Channel} with an extended SNR range from 0dB to 30dB and the maximum Doppler shift is from 0Hz to 97Hz. For the rest of this paper, we only train neural networks on the CE channel with $\zeta = \frac{1}{3f_{space}\left(N_f-1\right)} = 0.3129$ms as the proposed design criteria has been validated. This ensures that the trained neural networks generalize to the channels \textbf{including but not just} those channels, which are just example channels to show the generalization results. 
\begin{figure*}
    \centering
    \subfloat[The MSE results of the trained InterpolateNet \label{sub1}]{%
        \includegraphics[width=0.33\textwidth]{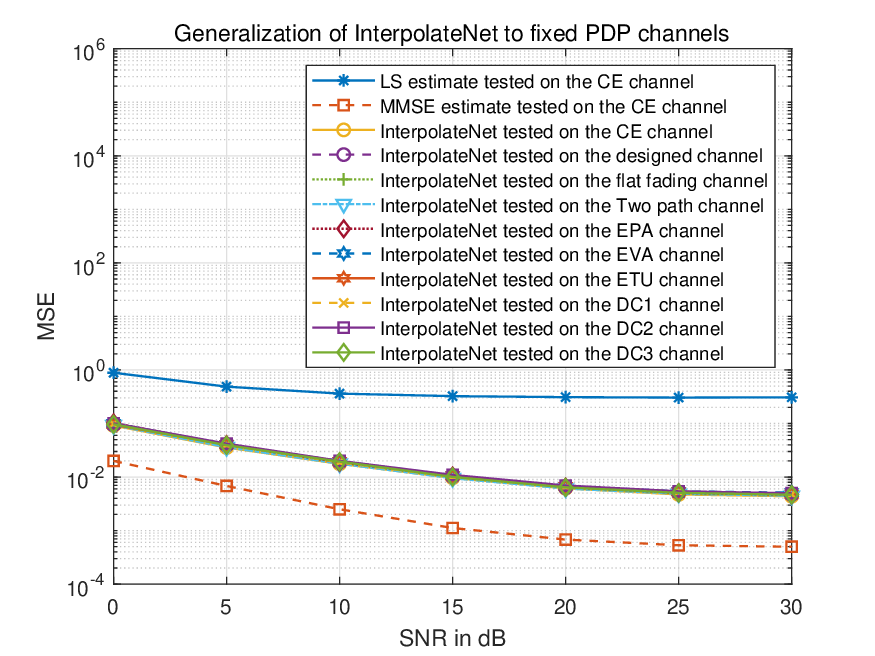}}
    \hfill
    \subfloat[The MSE results of the trained Channelformer \label{sub3}]{%
        \includegraphics[width=0.33\textwidth]{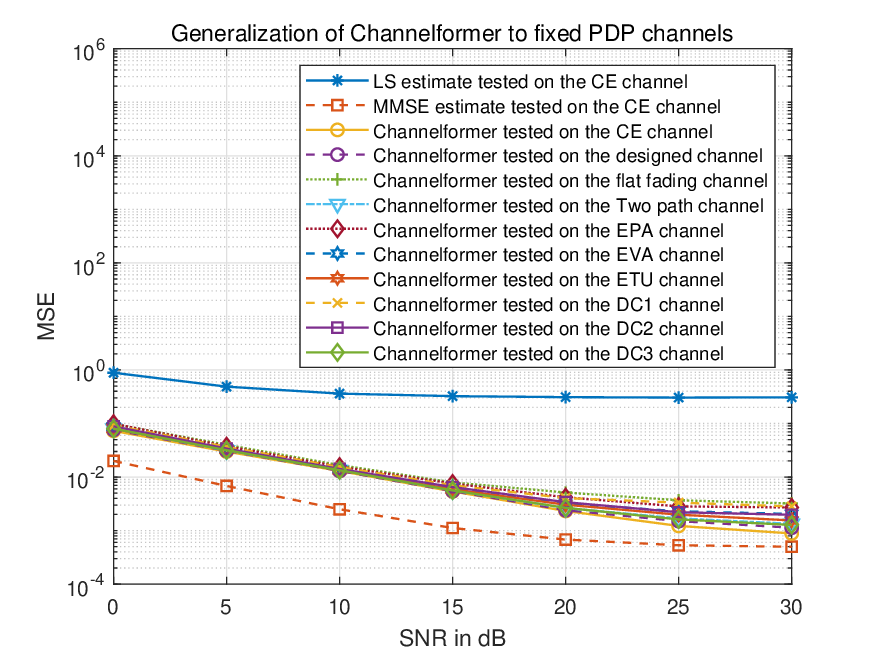}}
    \subfloat[The MSE results of the trained SimpleNet \label{sub5}]{%
       \includegraphics[width=0.33\textwidth]{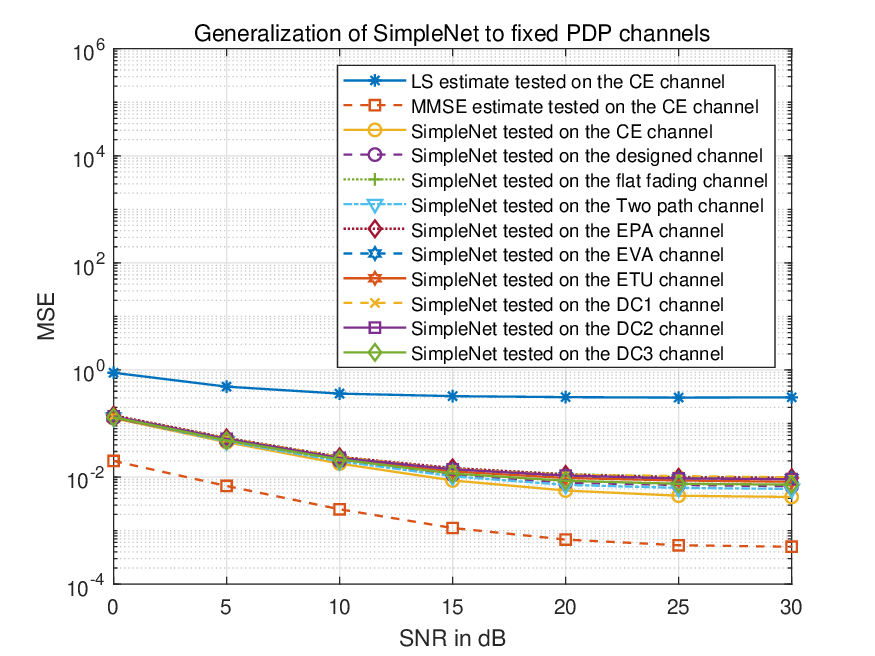}}
    \hfill
    \\
    \subfloat[The BER results of the trained InterpolateNet \label{sub2}]{%
        \includegraphics[width=0.33\textwidth]{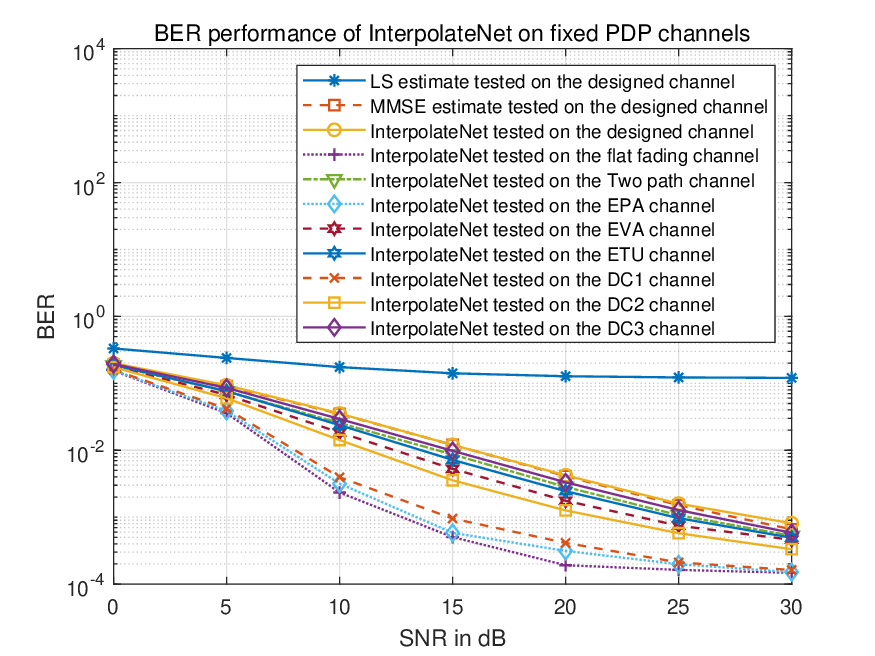}}
    \hfill
    \subfloat[The BER results of the trained Channelformer \label{sub4}]{%
        \includegraphics[width=0.33\textwidth]{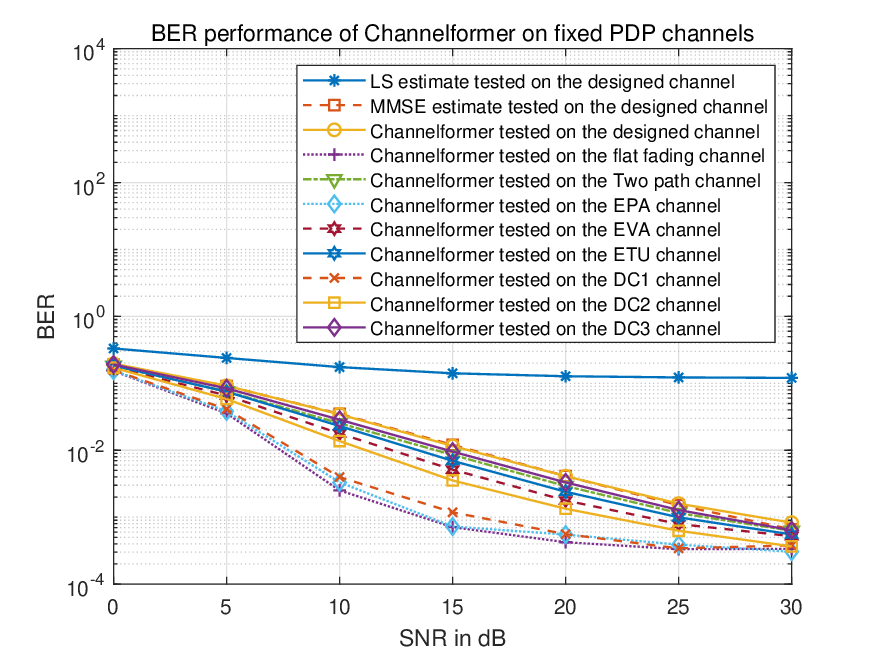}}
    \hfill
    \subfloat[The BER results of the trained SimpleNet \label{sub6}]{%
        \includegraphics[width=0.33\textwidth]{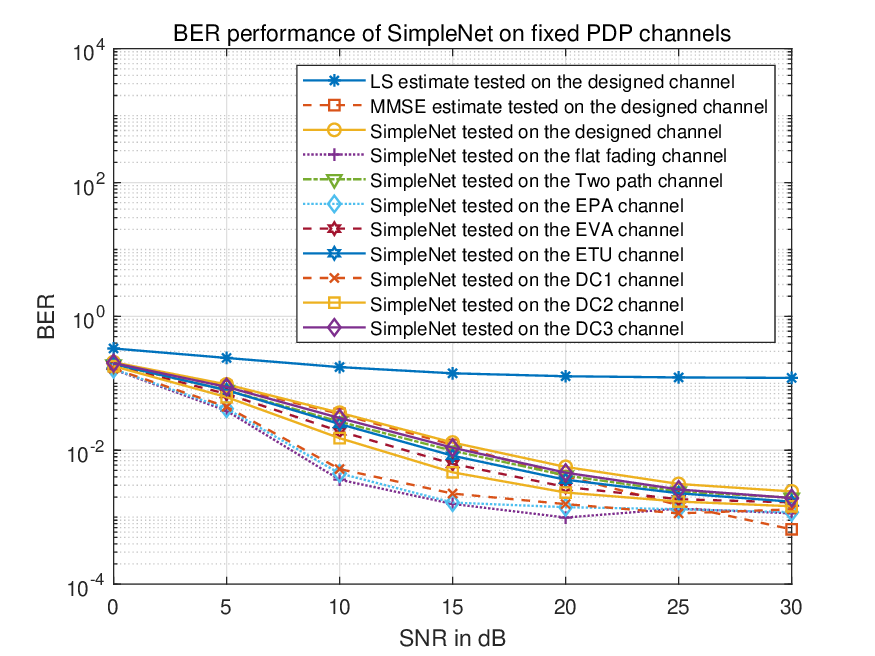}}
  \caption{MSE and BER results for the trained InterpolateNet, Channelformer and SimpleNet tested on the fixed PDP channels with an extended SNR range. The CE channel is used to generate the training dataset. }
  \label{sub-6}
\end{figure*}

Fig.~\ref{sub-6} provides both the MSE and BER results of the trained neural networks tested on the fixed PDP channels. The trained InterpolateNet maintains almost same MSE performance on all the fixed PDP channels in Fig.\ref{sub1}, while the trained SimpleNet has slight degradation of 0.0054 at the 30dB SNR in Fig.~\ref{sub3} because of the insufficient neural capacity (due to the small number of tunable parameters). As shown in Fig.~\ref{sub5}, the trained attention-based Channelformer has improved MSE performance but the fluctuations appear in high SNR range, where the MSE varies from 0.0009 to 0.0032 at 30 dB SNR. This is because the encoder pre-processes the input features to recognize the characteristics of the channel. At an SNR of 30dB, the MSE performance is very close to that of the MMSE method. However, the MSE performance gives an on-average evaluation which includes the channel gains at the pilot positions, i.e. $\mathbb{E}\left\{\left\Arrowvert\mathbf{\hat{H}} - \mathbf{H}\right\Arrowvert_{F}^{2}\right\}$. 

Therefore, we further investigate the BER results in Fig.~\ref{sub2}, \ref{sub4}, \ref{sub6}. BER results only depend on the estimate precision of the data symbols, to show the importance of the time interpolation as previously discussed. The BER variations at 30dB SNR are from 0.00012 to 0.00073 for InterpolateNet and from 0.00032 to 0.00088 for Channelformer respectively. SimpleNet has a BER variation from 0.0010 to 0.0020 at 30dB SNR, which is much worse than the InterpolateNet and Channelformer. The maximum BER variation for InterpolateNet, Channelformer and SimpleNet appears at 15 dB SNR with BER values from 0.000503 to 0.0147, from 0.000609 to 0.0145 and from 0.0015 to 0.0152. For several test channels, the trained InterpolateNet and Channelformer outperform the MMSE method in BER performance while the MMSE method has better MSE performance than neural networks solutions. The reason is, the deployed MMSE method exploits linear interpolation to estimate the channel matrix at the data symbols, which is imprecise. By being trained with the complete and actual channel matrix, the neural networks are able to predict the complete channel matrix which achieves better time interpolation than conventional methods. Meanwhile, the denoising capability is also retained by exploiting actual $\mathbf{H}$ directly rather than generating the estimated labels. To retain the denoising and interpolation capability, it is essential to train the network with the exact and complete channel matrix $\mathbf{H}$. As both the PDP and the SNR range of the channel designed for training are known, the neural network performance is deterministic as long as the environment SNR and delay spread are precisely measured following \cite{arslan2003noise} \cite{arslan2003delay}. 

In order to straightforwardly measure the precision at the data symbols, the SNR difference ($\Delta$SNR as loss) to achieve 1\% and 0.1\% BER among Fig.~\ref{BER}, Fig.~\ref{sub2}, Fig.~\ref{sub4} and Fig.~\ref{sub6}. Compared with Fig.~\ref{BER}, the trained InterpolateNet and Channelformer has only loss of 1.5dB maximum over these channels to achieve BER of 1\%, while the maximum loss for SimpleNet is 1.9dB. Thus, by using the $\Delta$ SNR, we observe results that are more consistent with Fig. 6 for InterpolateNet and Channelformer. For InterpolateNet, the maximum loss over these channels is only 2.35dB to achieve 0.1\% BER, compared with Fig.~\ref{BER}. The loss variation for Channelformer is relatively drastic because of the attention-based encoder. The maximum value of 3.8dB appears on the DC3 channel while the minimum value is 0.7 for the flat fading channel. The SimpleNet has extremely low complexity and performs more poorly than the other neural networks, resulting in not achieving 0.1\% BER for all channels. 
\begin{figure*}
    \centering
    \subfloat[$\Delta$SNR values for InterpolateNet]{%
        \includegraphics[width=0.33\textwidth]{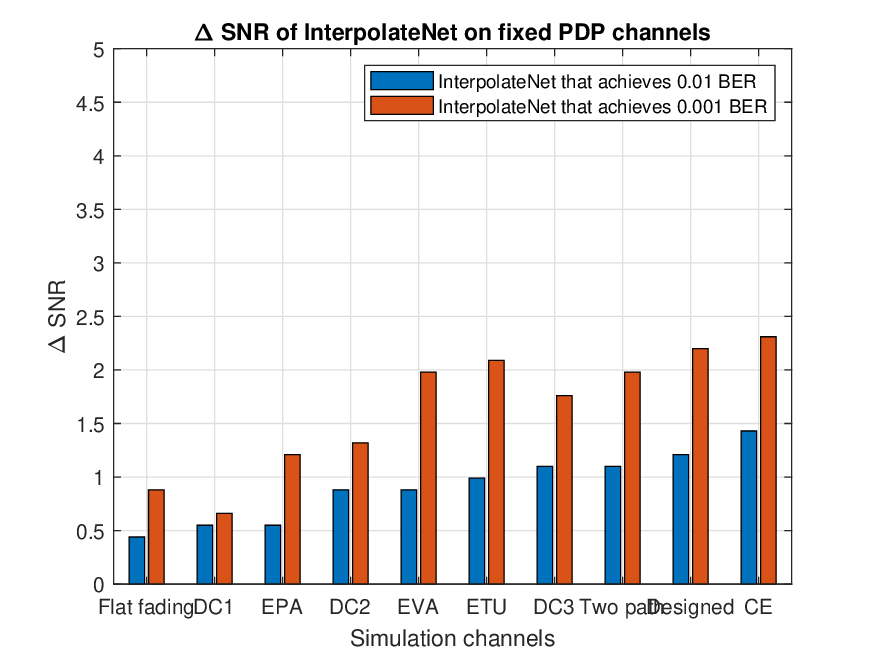}}
    \hfill
    \subfloat[$\Delta$SNR values for Channelformer]{%
        \includegraphics[width=0.33\textwidth]{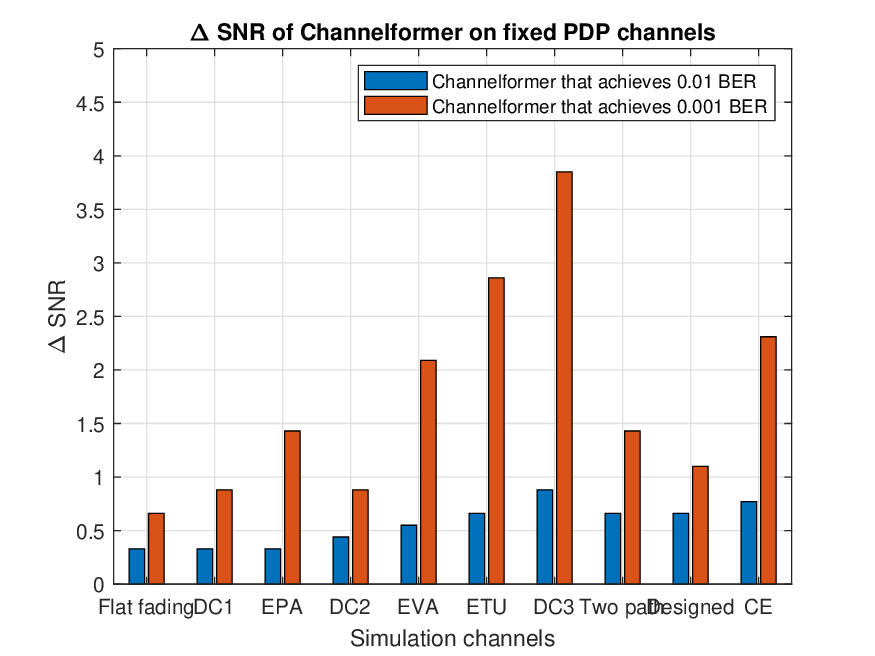}}
    \hfill
    \subfloat[$\Delta$SNR values for SimpleNet]{%
        \includegraphics[width=0.33\textwidth]{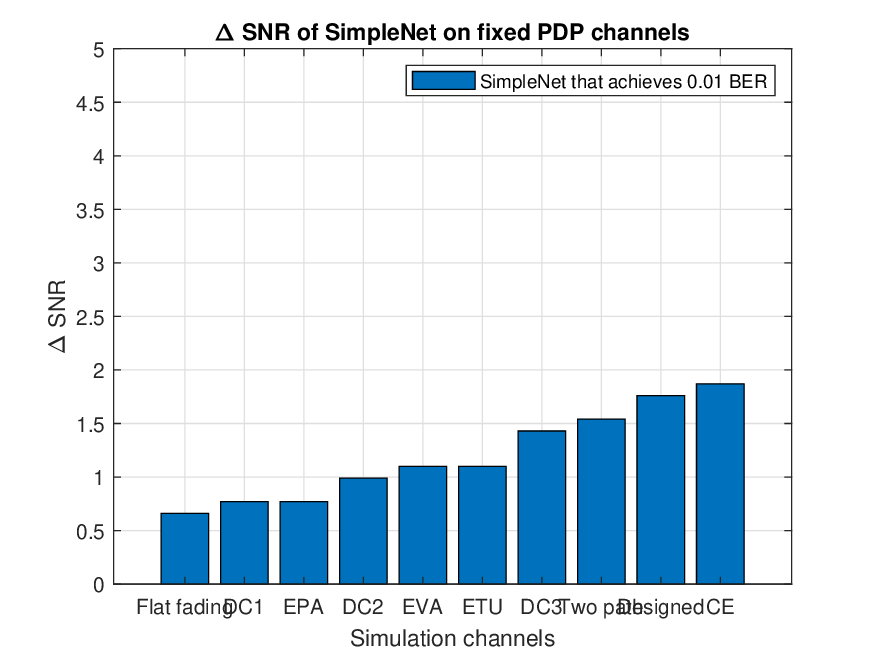}}
    \hfill
  \caption{$\Delta$SNR values for the trained InterpolateNet, Channelformer and SimpleNet tested on the fixed PDP channels. The CE channel is used to generate the training dataset}
  \label{sub-6 BER}
\end{figure*}
\subsection{Generalization to variable delay spread channels}
\label{Simulation_4}
The generalization achieved is also simulated on other channels with a new channel modeling. This is crucial because realistic channels may have different characteristics than existing channel modeling methods. Thus, this test would demonstrate the performance of the achieved generalization when operating in realistic environments. To validate the generalization to channels with this form of channel model, CDL-A, CDL-B, CDL-C, TDL-A and TDL-B channels are implemented with a very wide range of $\mathrm{DS_{desired}}$ values. 
\begin{figure*}
    \centering
    \subfloat[CE-trained InterpolateNet (extended SNR)\label{milli1}]{%
        \includegraphics[width=0.33\textwidth]{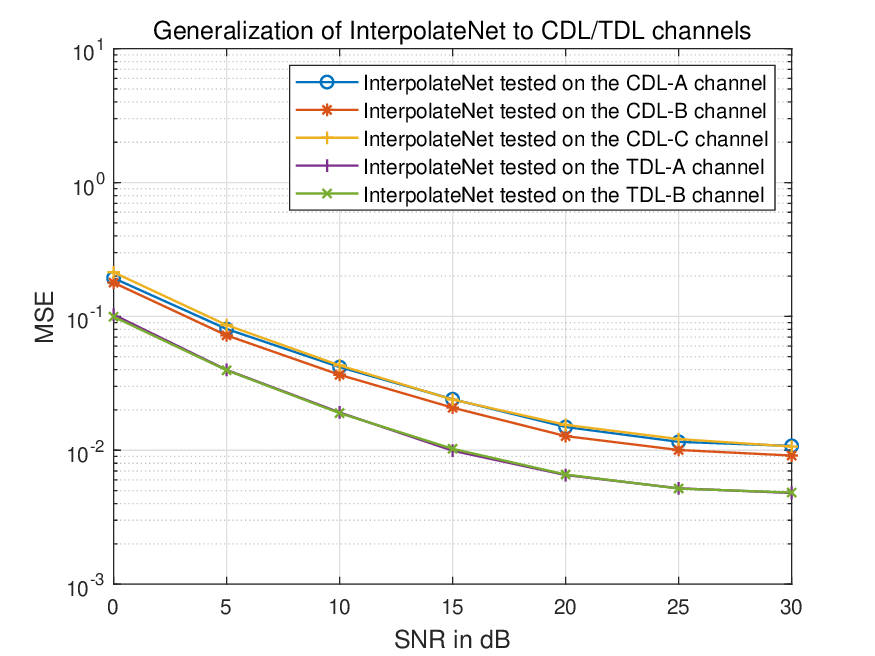}}
    \hfill
    \subfloat[CE-trained Channelformer (extended SNR)\label{milli3}]{%
        \includegraphics[width=0.33\textwidth]{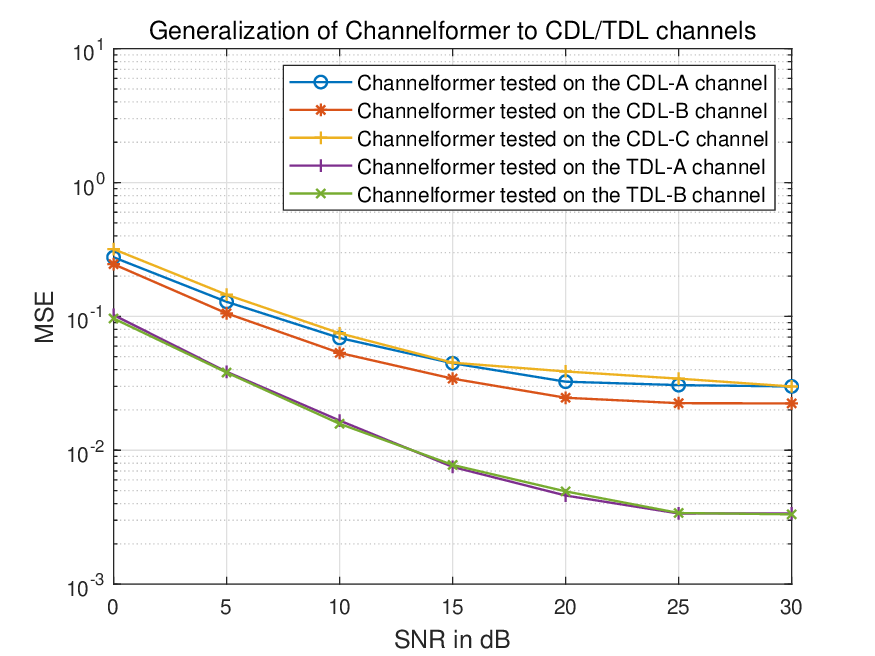}}
    \hfill
    \subfloat[CE-trained SimpleNet (extended SNR)\label{milli5}]{%
       \includegraphics[width=0.33\textwidth]{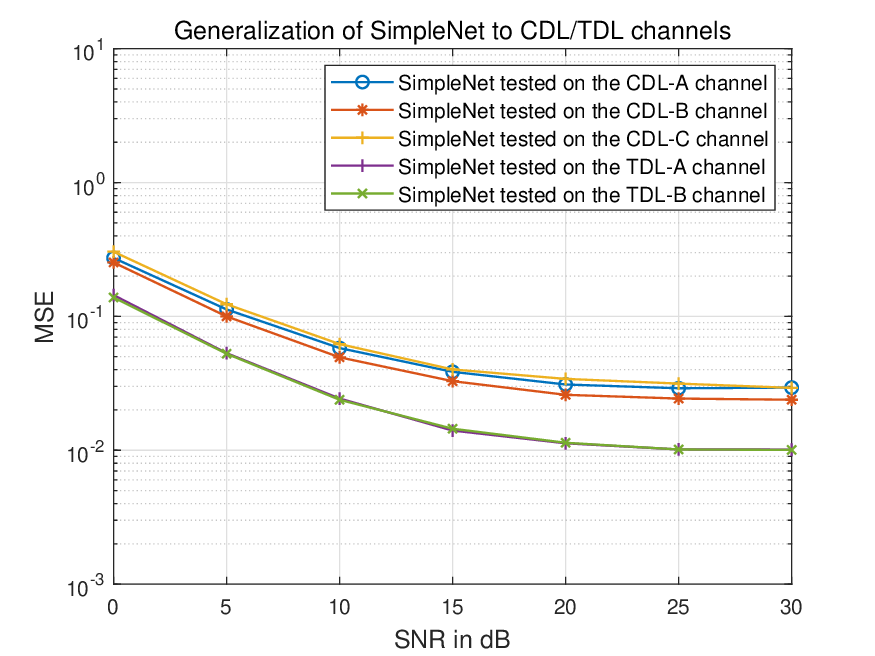}}
    \\
    \subfloat[CE-trained InterpolateNet (different $\mathrm{DS_{desired}}$)\label{milli2}]{%
        \includegraphics[width=0.33\textwidth]{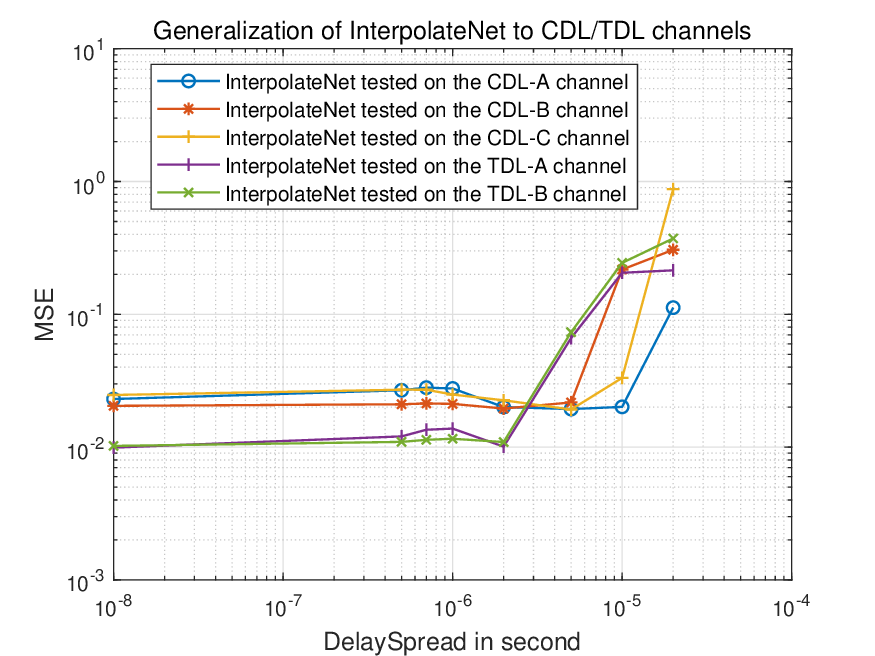}}
    \hfill
    \subfloat[CE-trained Channelformer (different $\mathrm{DS_{desired}}$)\label{milli4}]{%
        \includegraphics[width=0.33\textwidth]{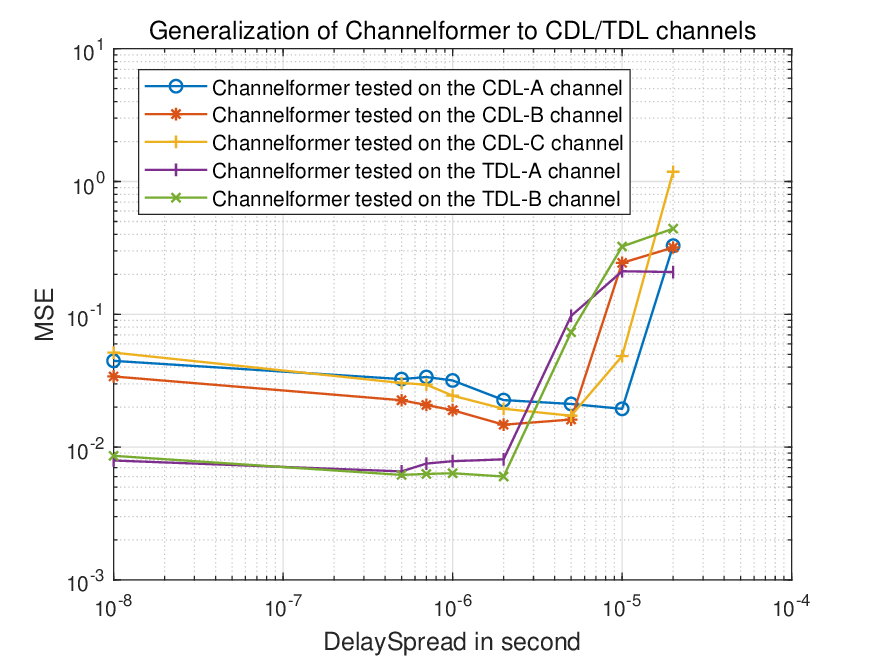}}
    \hfill
    \subfloat[CE-trained SimpleNet (different $\mathrm{DS_{desired}}$)\label{milli6}]{%
        \includegraphics[width=0.33\textwidth]{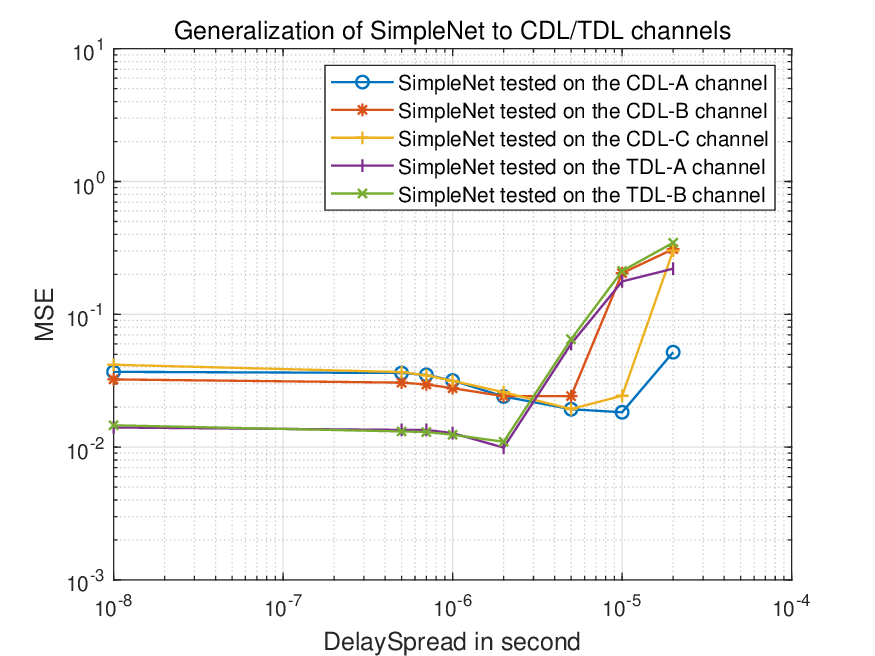}}
  \caption{MSE results for the CE-trained InterpolateNet, Channelformer and SimpleNet tested on the CDL/TDL channels operating at 39GHz with an extended SNR range at the left side and with a varied $\mathrm{DS_{desired}}$ at the right side. The channel used to generate the training dataset is the CE channel}
\end{figure*}

Fig.~\ref{milli1}, \ref{milli3}, \ref{milli5} provide the MSE results on the CDL/TDL channels with an extended SNR range from 0dB to 30dB and $\mathrm{DS_{desired}}$ = 30ns. Each trained neural network has nearly same MSE values when tested on the TDL-A and TDL-B channels, and the maximum MSE degradation for all the trained neural networks is 0.000032 at an SNR of 30dB. The MSE performance of the trained neural networks has a slight variation on the CDL-A, CDL-B and CDL-C channels, where the maximum MSE variation are 0.0015 for InterpolateNet, 0.0051 for SimpleNet and 0.0076 for Channelformer at an SNR of 30dB. This demonstrates that the achieved generalization to the CDL/TDL channels is still robust. Compared to Fig.~\ref{sub-6}, the MSE performance in Fig.~\ref{milli1} is slightly worse because of the channel modelling mismatches between the CDL/TDL channels and fixed PDP channels. 

Fig.~\ref{milli2}, \ref{milli4}, \ref{milli6} provide the MSE results on the CDL/TDL channels with variable $\mathrm{DS_{desired}}$ values from 10ns to 20,000ns and a fixed SNR of 15dB. When $\mathrm{DS_{desired}} \leq$ 1000ns, the trained InterpolateNet, Channelformer and SimpleNet maintain nearly same performance on the TDL channels and the CDL channels respectively. For the carrier frequency of 39GHz, the delay spread ranges from 16ns (Short-delay profile, Indoor office) to 786ns (Long-delay profile, Urban Macro) covering all the $\mathrm{DS_{desired}}$ values listed in 3GPP 38.901. Therefore, trained neural networks retain robust generalization to the CDL/TDL channels for the 3GPP current standards. For $\mathrm{DS_{desired}} \geq$ 1000ns, the MSE values generally increase with the growth of $\mathrm{DS_{desired}}$ because the maximum delay of the PDP starts to exceed the CP duration leading to inter-symbol interference (ISI). Therefore, all the MSE values at the sample of $\mathrm{DS_{desired}}$ = 20,000ns are very high. These results verify that the CE channel provides the benchmark training dataset for wireless channel estimation neural solutions to support general intelligence, and the MSE precision of the trained InterpolateNet, Channelformer and SimpleNet is shown to be sufficient for real-world implementation. 
\subsection{Adaption to different system hyper-parameters}
\label{Simulation_5}
To demonstrate the system scalability, we change some system hyper-parameters and then train the neural networks on the CE channel. The channel conditions for testing are identical with Section.~\ref{Simulation_sub6}. 
\begin{figure*}[htbp]
\centering
\subfloat[InterpolateNet (alternative DM-RS pattern) \label{Alternative_pattern_InterpolateNet}]{%
       \includegraphics[width=0.333\linewidth]{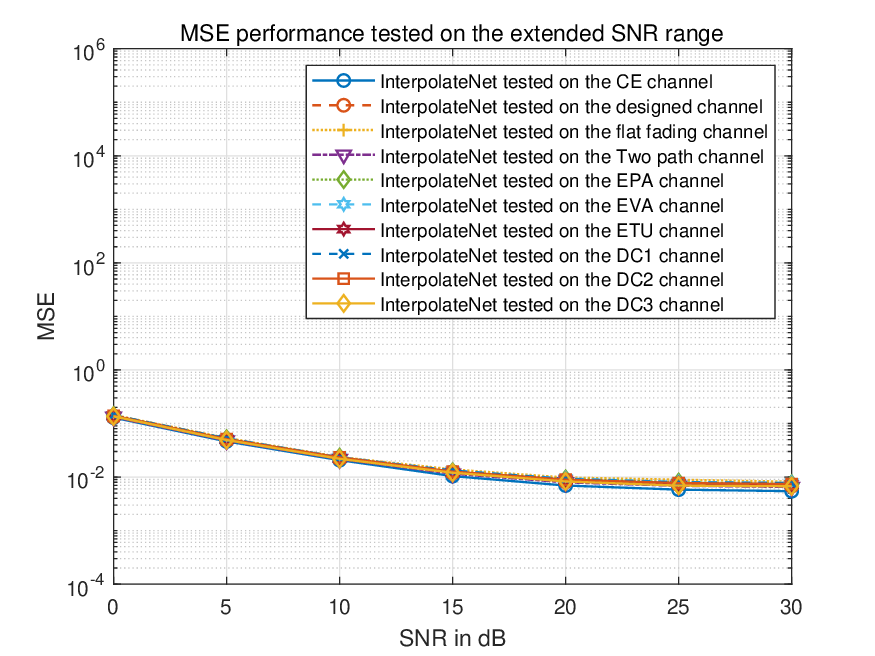}}
\hfill
\subfloat[Channelformer (alternative DM-RS pattern) \label{Alternative_pattern_Channelformer}]{%
        \includegraphics[width=0.333\linewidth]{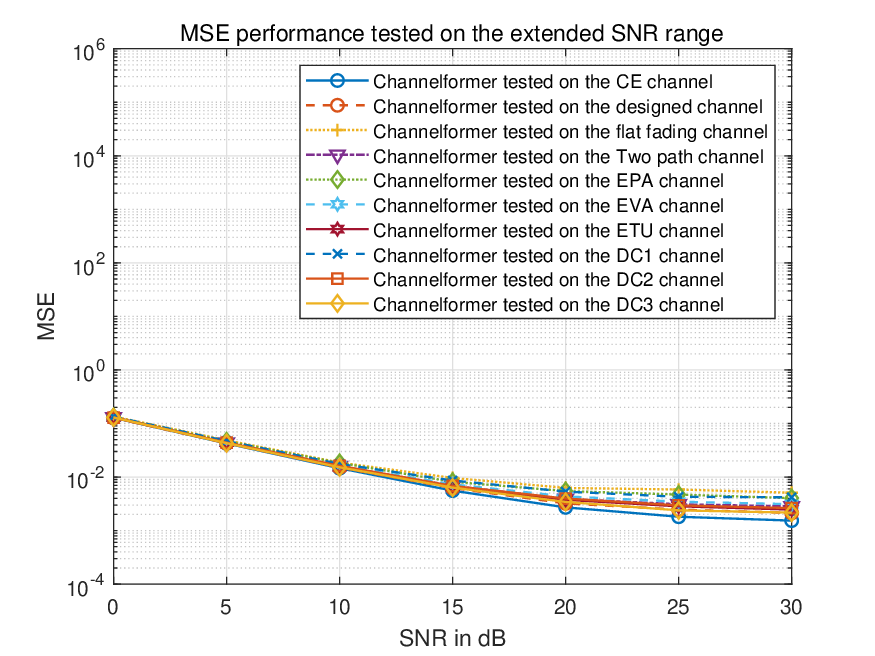}}
\hfill
\subfloat[SimpleNet (alternative DM-RS pattern) \label{Alternative_pattern_SimpleNet}]{%
       \includegraphics[width=0.333\linewidth]{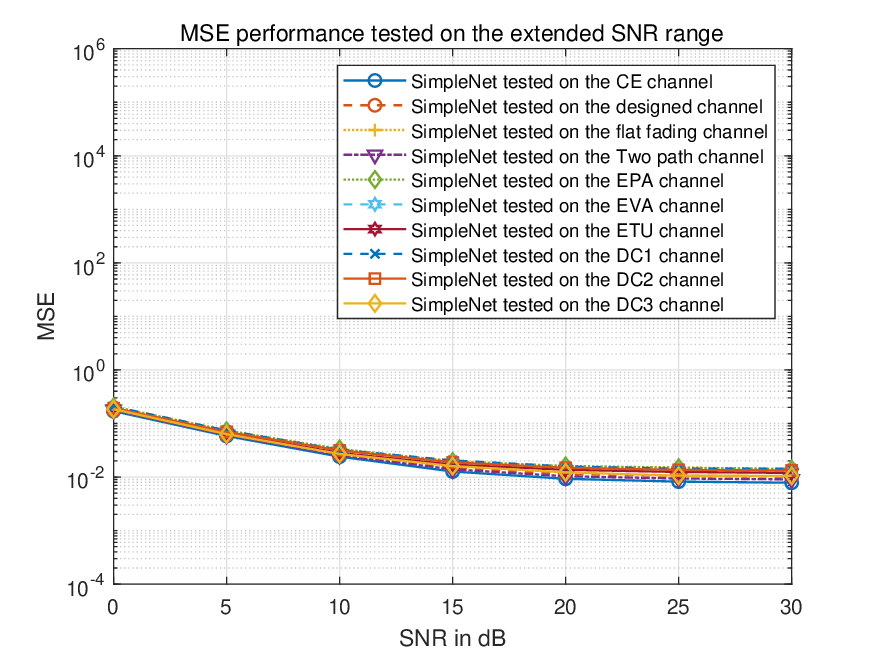}}
\hfill
\caption{MSE results for the neural networks tested on the alternative DM-RS pattern}
\end{figure*}
Fig.~\ref{Alternative_pattern_InterpolateNet}, \ref{Alternative_pattern_Channelformer}, \ref{Alternative_pattern_SimpleNet} provide the simulation results that train and test the neural networks with the alternative DM-RS pattern as shown in the right side of Fig.~\ref{DM-RS pattern}. The trained InterpolateNet has nearly the same MSE performance on each channel while the trained SimpleNet has a slight degradation in the high SNR range. Compared with InterpolateNet and SimpleNet, the trained Channelformer has relatively improved performance and slightly degraded generalization. Therefore, this design criteria is also valid for the alternative DM-RS pattern which demonstrates the general applicability of the proposed method. 

We train and test the neural networks with different $N_f = 24,36,72,96,120$ and batch sizes when tested on the fixed PDP channels with a fixed SNR of 15dB. For these $N_f$, the trained neural networks retain similar MSE performance that increases slightly with $N_f$, which increases from 0.0069 to 0.0117 (from $N_f = 24$ to $N_f = 120$) for InterpolateNet, from 0.0073 to 0.0120 for SimpleNet and from 0.0020 to 0.0037 for Channelformer. Therefore, good generalization is still achieved when the number of subcarriers $N_f$ is increased. For minibatch size choice during the training process, the MSE values of the trained neural networks are almost same among these channels (variation within 0.003), even with a small batch size of 16. Therefore, the achieved generalization is not significantly affected by the choice of batch size. 
\section{Conclusion}
\label{Conclusion}
We propose design criteria to generate the training data to achieve good generalization properties for channel estimation neural networks. The trained neural networks generalize robustly to new and previously unseen channels, and the proposed design criteria inherently determine the range of applicable channels. In this way, trained neural networks can estimate the complete channel matrix directly to retain interpolation capability, even on previously unseen channels. In addition, we propose a benchmark design which provides the maximum generalization to wireless channels. Moreover, the proposed design criteria is applicable for most neural networks because no specific neural architecture is required. From the simulation results, trained neural networks achieve a certain performance on all these fixed PDP channels, compared with the performance on the designed channel. Moreover, they generalize robustly to CDL/TDL channels with variable delay spreads, which have a different channel modelling compared with the channel designed for training. We also show that the MSE equation for the MMSE filter can be used to assess the generalization capability to validate the effectiveness of the proposed method. This paper addresses the problem that offline-trained neural networks cannot perform well in new environments, to avoid online training and retain superior performance of neural networks. This underpins the low-latency and low-complexity properties for real-world implementation of channel estimation neural networks. 
\section*{Acknowledgement}
For the purpose of open access, the authors have applied a Creative Commons Attribution (CC BY) licence to any Author Accepted Manuscript version arising from this submission. This research is supported by EPSRC projects EP/X04047X/2 and EP/Y037243/1. To support reproducibility, the simulation code can be downloaded at https://github.com/dianixn/Achieving-Robust-Channel-Estimation-Neural-Networks-by-Designed-Training-Data. 
\ifCLASSOPTIONcaptionsoff
  \newpage
\fi

\bibliographystyle{IEEEtran}

\bibliography{Reference}

\begin{thebibliography}{10}
\providecommand{\url}[1]{#1}
\csname url@samestyle\endcsname
\providecommand{\newblock}{\relax}
\providecommand{\bibinfo}[2]{#2}
\providecommand{\BIBentrySTDinterwordspacing}{\spaceskip=0pt\relax}
\providecommand{\BIBentryALTinterwordstretchfactor}{4}
\providecommand{\BIBentryALTinterwordspacing}{\spaceskip=\fontdimen2\font plus
\BIBentryALTinterwordstretchfactor\fontdimen3\font minus \fontdimen4\font\relax}
\providecommand{\BIBforeignlanguage}[2]{{%
\expandafter\ifx\csname l@#1\endcsname\relax
\typeout{** WARNING: IEEEtran.bst: No hyphenation pattern has been}%
\typeout{** loaded for the language `#1'. Using the pattern for}%
\typeout{** the default language instead.}%
\else
\language=\csname l@#1\endcsname
\fi
#2}}
\providecommand{\BIBdecl}{\relax}
\BIBdecl

\bibitem{wang2023road}
C.-X. Wang \emph{et~al.}, ``On the road to 6{G}: Visions, requirements, key technologies and testbeds,'' \emph{IEEE Commun. Surveys Tuts.}, 2023.

\bibitem{wu2024intelligent}
Q.~Wu, B.~Zheng, C.~You, L.~Zhu, K.~Shen, X.~Shao, W.~Mei, B.~Di, H.~Zhang, E.~Basar \emph{et~al.}, ``Intelligent surfaces empowered wireless network: Recent advances and the road to 6g,'' \emph{Proceedings of the IEEE}, 2024.

\bibitem{van1995channel}
J.-J. Van De~Beek, O.~Edfors, M.~Sandell, S.~K. Wilson, and P.~O. Borjesson, ``On channel estimation in {OFDM} systems,'' in \emph{1995 IEEE 45th VTC.}, vol.~2.\hskip 1em plus 0.5em minus 0.4em\relax IEEE, 1995, pp. 815--819.

\bibitem{edfors1998ofdm}
O.~Edfors, M.~Sandell, J.-J. Van~de Beek, S.~K. Wilson, and P.~O. Borjesson, ``{OFDM} channel estimation by singular value decomposition,'' \emph{IEEE Trans. Commun.}, vol.~46, no.~7, pp. 931--939, 1998.

\bibitem{deng2003decision}
X.~Deng, A.~M. Haimovich, and J.~Garcia-Frias, ``Decision directed iterative channel estimation for {MIMO} systems,'' in \emph{IEEE ICC'03.}, vol.~4.\hskip 1em plus 0.5em minus 0.4em\relax IEEE, 2003, pp. 2326--2329.

\bibitem{orozco2004channel}
A.~G. Orozco-Lugo, M.~M. Lara, and D.~C. McLernon, ``Channel estimation using implicit training,'' \emph{IEEE Transactions on Signal Processing}, vol.~52, no.~1, pp. 240--254, 2004.

\bibitem{neumann2018learning}
D.~Neumann, T.~Wiese, and W.~Utschick, ``Learning the {MMSE} channel estimator,'' \emph{IEEE Transactions on Signal Processing}, vol.~66, no.~11, pp. 2905--2917, 2018.

\bibitem{koller2022asymptotically}
M.~Koller, B.~Fesl, N.~Turan, and W.~Utschick, ``An asymptotically mse-optimal estimator based on {Gaussian} mixture models,'' \emph{IEEE Transactions on Signal Processing}, vol.~70, pp. 4109--4123, 2022.

\bibitem{baur2024channel}
M.~Baur, N.~Turan, B.~Fesl, and W.~Utschick, ``Channel estimation in underdetermined systems utilizing variational autoencoders,'' in \emph{ICASSP 2024-2024 IEEE International Conference on Acoustics, Speech and Signal Processing (ICASSP)}.\hskip 1em plus 0.5em minus 0.4em\relax IEEE, 2024, pp. 9031--9035.

\bibitem{baur2024leveraging}
M.~Baur, B.~Fesl, and W.~Utschick, ``Leveraging variational autoencoders for parameterized {MMSE} estimation,'' \emph{IEEE Transactions on Signal Processing}, 2024.

\bibitem{luan2025robust}
D.~Luan and J.~Thompson, ``Robust channel estimation for optical wireless communications using neural network,'' \emph{arXiv preprint arXiv:2504.02134}, 2025.

\bibitem{soltani2019deep}
M.~Soltani, V.~Pourahmadi, A.~Mirzaei, and H.~Sheikhzadeh, ``Deep learning-based channel estimation,'' \emph{IEEE Communications Letters}, vol.~23, no.~4, pp. 652--655, 2019.

\bibitem{bahdanau2014neural}
D.~Bahdanau \emph{et~al.}, ``Neural machine translation by jointly learning to align and translate,'' \emph{arXiv preprint arXiv:1409.0473}, 2014.

\bibitem{mashhadi2021pruning}
M.~B. Mashhadi and D.~G{\"u}nd{\"u}z, ``Pruning the pilots: Deep learning-based pilot design and channel estimation for {MIMO-OFDM} systems,'' \emph{IEEE Trans. Wireless Commun.}, vol.~20, no.~10, pp. 6315--6328, 2021.

\bibitem{gao2021attention}
J.~Gao, M.~Hu, C.~Zhong, G.~Y. Li, and Z.~Zhang, ``An attention-aided deep learning framework for massive {MIMO} channel estimation,'' \emph{IEEE Trans. Wireless Commun.}, 2021.

\bibitem{luan2022attention}
D.~Luan and J.~Thompson, ``Attention based neural networks for wireless channel estimation,'' in \emph{2022 IEEE 95th Vehicular Technology Conference:(VTC2022-Spring)}.\hskip 1em plus 0.5em minus 0.4em\relax IEEE, 2022, pp. 1--5.

\bibitem{luan2023channelformer}
D.~Luan and J.~S. Thompson, ``Channelformer: Attention based neural solution for wireless channel estimation and effective online training,'' \emph{IEEE Transactions on Wireless Communications}, vol.~22, no.~10, pp. 6562--6577, 2023.

\bibitem{vaswani2017attention}
A.~Vaswani, N.~Shazeer, N.~Parmar, J.~Uszkoreit, L.~Jones, A.~N. Gomez, {\L}.~Kaiser, and I.~Polosukhin, ``Attention is all you need,'' \emph{Advances in neural information processing systems}, vol.~30, 2017.

\bibitem{zheng2021online}
X.~Zheng and V.~K. Lau, ``Online deep neural networks for mmwave massive mimo channel estimation with arbitrary array geometry,'' \emph{IEEE Transactions on Signal Processing}, vol.~69, pp. 2010--2025, 2021.

\bibitem{mohajer2024dynamic}
A.~Mohajer, J.~Hajipour, and V.~C. Leung, ``Dynamic offloading in mobile edge computing with traffic-aware network slicing and adaptive td3 strategy,'' \emph{IEEE Communications Letters}, 2024.

\bibitem{minn2000investigation}
H.~Minn and V.~K. Bhargava, ``An investigation into time-domain approach for {OFDM} channel estimation,'' \emph{IEEE Transactions on broadcasting}, vol.~46, no.~4, pp. 240--248, 2000.

\bibitem{tang2007pilot}
Z.~Tang, R.~C. Cannizzaro, G.~Leus, and P.~Banelli, ``Pilot-assisted time-varying channel estimation for {OFDM} systems,'' \emph{IEEE Transactions on Signal Processing}, vol.~55, no.~5, pp. 2226--2238, 2007.

\bibitem{nissel2018doubly}
R.~Nissel, F.~Ademaj, and M.~Rupp, ``Doubly-selective channel estimation in {FBMC-OQAM} and {OFDM} systems,'' in \emph{2018 IEEE 88th Vehicular Technology Conference (VTC-Fall)}.\hskip 1em plus 0.5em minus 0.4em\relax IEEE, 2018, pp. 1--5.

\bibitem{borah1999frequency}
D.~K. Borah and B.~Hart, ``Frequency-selective fading channel estimation with a polynomial time-varying channel model,'' \emph{IEEE Trans. Commun.}, vol.~47, no.~6, pp. 862--873, 1999.

\bibitem{mccloskey1989catastrophic}
M.~McCloskey and N.~J. Cohen, ``Catastrophic interference in connectionist networks: The sequential learning problem,'' in \emph{Psychology of learning and motivation}.\hskip 1em plus 0.5em minus 0.4em\relax Elsevier, 1989, vol.~24, pp. 109--165.

\bibitem{bousquet2002stability}
O.~Bousquet and A.~Elisseeff, ``Stability and generalization,'' \emph{The Journal of Machine Learning Research}, vol.~2, pp. 499--526, 2002.

\bibitem{zhang2021understanding}
C.~Zhang, S.~Bengio, M.~Hardt, B.~Recht, and O.~Vinyals, ``Understanding deep learning (still) requires rethinking generalization,'' \emph{Communications of the ACM}, vol.~64, no.~3, pp. 107--115, 2021.

\bibitem{advani2020high}
M.~S. Advani, A.~M. Saxe, and H.~Sompolinsky, ``High-dimensional dynamics of generalization error in neural networks,'' \emph{Neural Networks}, vol. 132, pp. 428--446, 2020.

\bibitem{akrout2023domain}
M.~Akrout, A.~Feriani, F.~Bellili, A.~Mezghani, and E.~Hossain, ``Domain generalization in machine learning models for wireless communications: Concepts, state-of-the-art, and open issues,'' \emph{IEEE Communications Surveys \& Tutorials}, 2023.

\bibitem{wang2020high}
H.~Wang, X.~Wu, Z.~Huang, and E.~P. Xing, ``High-frequency component helps explain the generalization of convolutional neural networks,'' in \emph{Proceedings of the IEEE/CVF CVPR}, 2020, pp. 8684--8694.

\bibitem{cavers1991analysis}
J.~K. Cavers, ``An analysis of pilot symbol assisted modulation for {Rayleigh} fading channels (mobile radio),'' \emph{IEEE transactions on vehicular technology}, vol.~40, no.~4, pp. 686--693, 1991.

\bibitem{garcia2006support}
M.-G. Garcia, J.~L. Rojo-{\'A}lvarez, F.~Alonso-Atienza, and M.~Mart{\'\i}nez-Ram{\'o}n, ``Support vector machines for robust channel estimation in {OFDM},'' \emph{IEEE Signal Processing Letters}, vol.~13, no.~7, pp. 397--400, 2006.

\bibitem{cai2004robust}
J.~Cai, X.~Shen, and J.~W. Mark, ``Robust channel estimation for {OFDM} wireless communication systems-an h/sub/spl infin//approach,'' \emph{IEEE Transactions on Wireless Communications}, vol.~3, no.~6, pp. 2060--2071, 2004.

\bibitem{demir2024efficient}
{\"O}.~T. Demir, E.~Bj{\"o}rnson, and L.~Sanguinetti, ``Efficient channel estimation with shorter pilots in ris-aided communications: Using array geometries and interference statistics,'' \emph{IEEE Transactions on Wireless Communications}, 2024.

\bibitem{10279223}
D.~Luan and J.~Thompson, ``Achieving robust generalization for wireless channel estimation neural networks by designed training data,'' in \emph{IEEE International Conference on Communications}, 2023, pp. 3462--3467.

\bibitem{patzold2009two}
M.~Patzold, C.-X. Wang, and B.~O. Hogstad, ``Two new sum-of-sinusoids-based methods for the efficient generation of multiple uncorrelated {Rayleigh} fading waveforms,'' \emph{IEEE Trans. Wireless Commun.}, vol.~8, no.~6, pp. 3122--3131, 2009.

\bibitem{luan2021low}
D.~Luan and J.~Thompson, ``Low complexity channel estimation with neural network solutions,'' in \emph{WSA 2021; 25th International ITG Workshop on Smart Antennas}.\hskip 1em plus 0.5em minus 0.4em\relax VDE, 2021, pp. 1--6.

\bibitem{kay1993fundamentals}
S.~M. Kay, \emph{Fundamentals of statistical signal processing: estimation theory}.\hskip 1em plus 0.5em minus 0.4em\relax Prentice-Hall, Inc., 1993.

\bibitem{li1998robust}
Y.~Li, L.~J. Cimini, and N.~R. Sollenberger, ``Robust channel estimation for {OFDM} systems with rapid dispersive fading channels,'' \emph{IEEE Transactions on communications}, vol.~46, no.~7, pp. 902--915, 1998.

\bibitem{srivastava2004robust}
V.~Srivastava, C.~K. Ho, P.~H.~W. Fung, and S.~Sun, ``Robust {MMSE} channel estimation in {OFDM} systems with practical timing synchronization,'' in \emph{2004 IEEE Wireless Communications and Networking Conference}, vol.~2.\hskip 1em plus 0.5em minus 0.4em\relax IEEE, 2004, pp. 711--716.

\bibitem{dent1993jakes}
P.~Dent, G.~E. Bottomley, and T.~Croft, ``Jakes fading model revisited,'' \emph{Electronics letters}, vol.~13, no.~29, pp. 1162--1163, 1993.

\bibitem{arslan2003noise}
H.~Arslan and S.~Reddy, ``Noise power and snr estimation for {OFDM} based wireless communication systems,'' in \emph{Proc. of 3rd IASTED International Conference on Wireless and Optical Communications (WOC), Banff, Alberta, Canada}, 2003.

\bibitem{arslan2003delay}
H.~Arslan and T.~Yucek, ``Delay spread estimation for wireless communication systems,'' in \emph{IEEE ISCC 2003}.\hskip 1em plus 0.5em minus 0.4em\relax IEEE, 2003, pp. 282--287.

\end{thebibliography}

\end{document}